\documentclass{aa}  
\usepackage{graphicx}
\usepackage{txfonts}
\usepackage{dirtytalk}
\usepackage[normalem]{ulem}
\usepackage{hyperref}

\usepackage{color}

\begin{document} 

   \title{Superflares on the late-type giant KIC\,2852961}
  \subtitle{Scaling effect behind flaring at different energy levels}

   \author{Zs.~K\H{o}v\'ari
          \inst{1}
        \and
          K.~Ol\'ah
          \inst{1}
        \and
          M.~N.~G\"unther
          \inst{2,}\thanks{Juan Carlos Torres Fellow}
        \and
          K. Vida
          \inst{1,3}
        \and
          L. Kriskovics
          \inst{1,3}
        \and
          B. Seli
          \inst{1}
        \and
          G.~\'A.~Bakos
          \inst{4}
        \and
        J.~D.~Hartman
          \inst{4}
        \and
        Z.~Csubry
          \inst{4}
        \and
          W.~Bhatti
          \inst{4}
  }

   \institute{Konkoly Observatory, Research Centre for Astronomy and Earth Sciences, Budapest, Hungary\\
              \email{kovari@konkoly.hu}
         \and
             Department of Physics, and Kavli Institute for Astrophysics and Space Research, Massachusetts Institute of Technology, Cambridge, MA 02139, USA
        \and
            ELTE Eötvös Loránd University, Institute of Physics, Budapest, Hungary
        \and
            Department of Astrophysical Sciences, Princeton University, NJ 08544, USA
            }

   \date{Received ...; accepted ...}

  \abstract
   {The most powerful superflares reaching $10^{39}$\,erg bolometric energy are from giant stars. The mechanism behind flaring is supposed to be the magnetic reconnection, which is closely related to magnetic activity including starspots. However, it is poorly understood, how the underlying magnetic dynamo works and how the flare activity is related to the stellar properties which eventually control the dynamo action.}
   {We analyse the flaring activity of KIC\,2852961, a late-type giant star, in order to understand how the flare statistics are related to that of other stars with flares and superflares and what the role of the observed stellar properties in generating flares is.}
   {We search for flares in the full \emph{Kepler} dataset of KIC\,2852961 by an automated technique together with visual inspection. We cross-match the flare-like events detected by the two different approaches and set a final list of 59 verified flares during the observing term. We calculate flare energies for the sample and perform a statistical analysis.}
   {The stellar properties of KIC\,2852961 are revised and a more consistent set of parameters are proposed. The cumulative flare energy distribution can be characterized by a broken power-law, i.e. on the log-log representation the distribution function is fitted by two linear functions with different slopes, depending on the energy range fitted. We find that the total flare energy integrated over a few rotation periods correlates with the average amplitude of the rotational modulation due to starspots. }
   {Flares and superflares seem to be the result of the same physical mechanism at different energetic levels, also implying that late-type stars in the main sequence and flaring giant stars have the same underlying physical process for emitting flares. There might be a scaling effect behind generating flares and superflares in the sense that the higher the magnetic activity the higher the overall magnetic energy released by flares and/or superflares.}

   \keywords{Stars: activity --
   Stars: flare --
   Stars: late-type --
   Stars: individual: KIC\,2852961, 2MASS J19261136+3803107, TIC\,137220334
               }
   \maketitle
%

\section{Introduction}\label{intro}

Studying the cosmic neighborhood of magnetically active stars, i.e., the impact of stellar magnetism on the circumstellar environment, where planets may revolve, is currently a hot issue. Stellar flares can heavily affect their close vicinity, such like the solar flares affect the Earth. The most energetic solar flares recorded so far, e.g. the \say{Carrington Event} in 1859, reached the energy output of $10^{33}$\,erg \citep{2013JSWSC...3A..31C}. On the other hand, stellar flares can release one to six orders of magnitude more energy \citep{2012Natur.485..478M} compared to the most powerful X-class solar flares; such `superflare stars' are mostly among solar-like stars from the main sequence, but can also be evolved stars in some measure, being either single or member of a binary system \citep[see, e.g.][and their references]{2015MNRAS.447.2714B,2018ARep...62..513K,2019ApJ...876...58N}.

The high magnetic energy outbursts by flares supposedly originate from magnetic reconnection, which presumes an underlying dynamo action, i.e. rotation/differential rotation interfering with convective motions. However, through stellar evolution slower rotation and increased size are expected to result in weaker magnetic fields and therefore lower level of magnetic activity for evolved stars compared with their main sequence progenitors. Yet, the most powerful superflares are from giants \citep{2015MNRAS.447.2714B}. Just recently, cross-matching superflare stars from the \emph{Kepler} catalogue with the \emph{Gaia} DR-2 stellar radius estimates has shown that more than 40\% of the previously supposed solar-type flare stars were subgiants \citep{2019ApJ...876...58N}. Magnetic activity is present and can indeed be strong along the red giant branch, which has been verified by direct imaging of starspots on the K-giant $\zeta$ Andromedae \citep{2016Natur.533..217R}. However, it is not clear what kind of mechanism could generate sufficient energy to provide the most powerful superflares on giant stars. It is quite certain that in some cases binarity plays a key role: a close companion star or a close-in giant planet could mediate magnetic reconnection and so provoke superflares \citep{1998A&A...335..248F,2000ApJ...529.1031R}. \citet{2018ARep...62..513K} proposed a magnetic dynamo working with antisolar differential rotation to explain the production of the most powerful superflares on giant stars. This is supported by the finding that only giant stars were reported so far to exhibit antisolar differential rotation \citep[see][and their references]{2017AN....338..903K}. But beside the non-uniform rotation profile, convective turbulence should also play a crucial role in driving stellar dynamos. Just recently, \citet{2020NatAs.tmp...46L} have demonstrated that a common dynamo scaling can be achieved for late-type main sequence and evolved, post-main-sequence stars only when both stellar rotation and convection are taken into account. This finding infers that magnetic dynamo action related flares in solar-type stars and superflares, for instance, in late-type giants can be linked by scaling as well.
The paradigm that dynamo action is necessary to produce flares, however, is nuanced by the recent finding that A-type stars without convective bulk can also have superflares \citep{2015MNRAS.447.2714B}. On the other hand, this finding is questioned by \citet{2017MNRAS.466.3060P} concluding that most of the A-type flare star candidates in \citet{2015MNRAS.447.2714B} are binary stars and flares probably originate from an unresolved companion star.

Observing flares in giant stars from the ground is quite a challenge, first of all, because of the luminous background of the stellar surface. The most energetic flares of $10^{38}$\,erg in the optical range would rise the brightness level of a red giant by only a few hundredth magnitude at the peak. Such a small change is in the order of the brightness variability due to short-term redistribution of starspots, i.e., generally flares in spotted giant stars can easily be indistinguishable from a small change in the rotational modulation in case of low data sampling and/or low data quality. At the same time, signals of less luminous flares would easily blend into the noise. Therefore, high precision space photometry from \emph{Kepler} or \emph{TESS} can be very useful in studying the optical signs of stellar activity in detail, including starspots and (super)flares, but it can also reveal such phenomena which are hardly or not observable from the ground, e.g. oscillations. Moreover, space instruments can provide continuous observations \citep[e.g.][]{1984ApJ...282..733B,2001ApJ...562L..83A,2002ApJ...570..799S,2006ApJS..164..173M} which are inevitably required to examine the temporal behaviour of complex (multiple) flare events or make up flare statistics for individual targets \citep[see also][etc.]{2016ApJ...829...23D,2017ApJ...841..124V,2019ApJ...884..160V,2020AJ....159...60G}.

\begin{figure}
   \centering
   \includegraphics[width=\columnwidth]{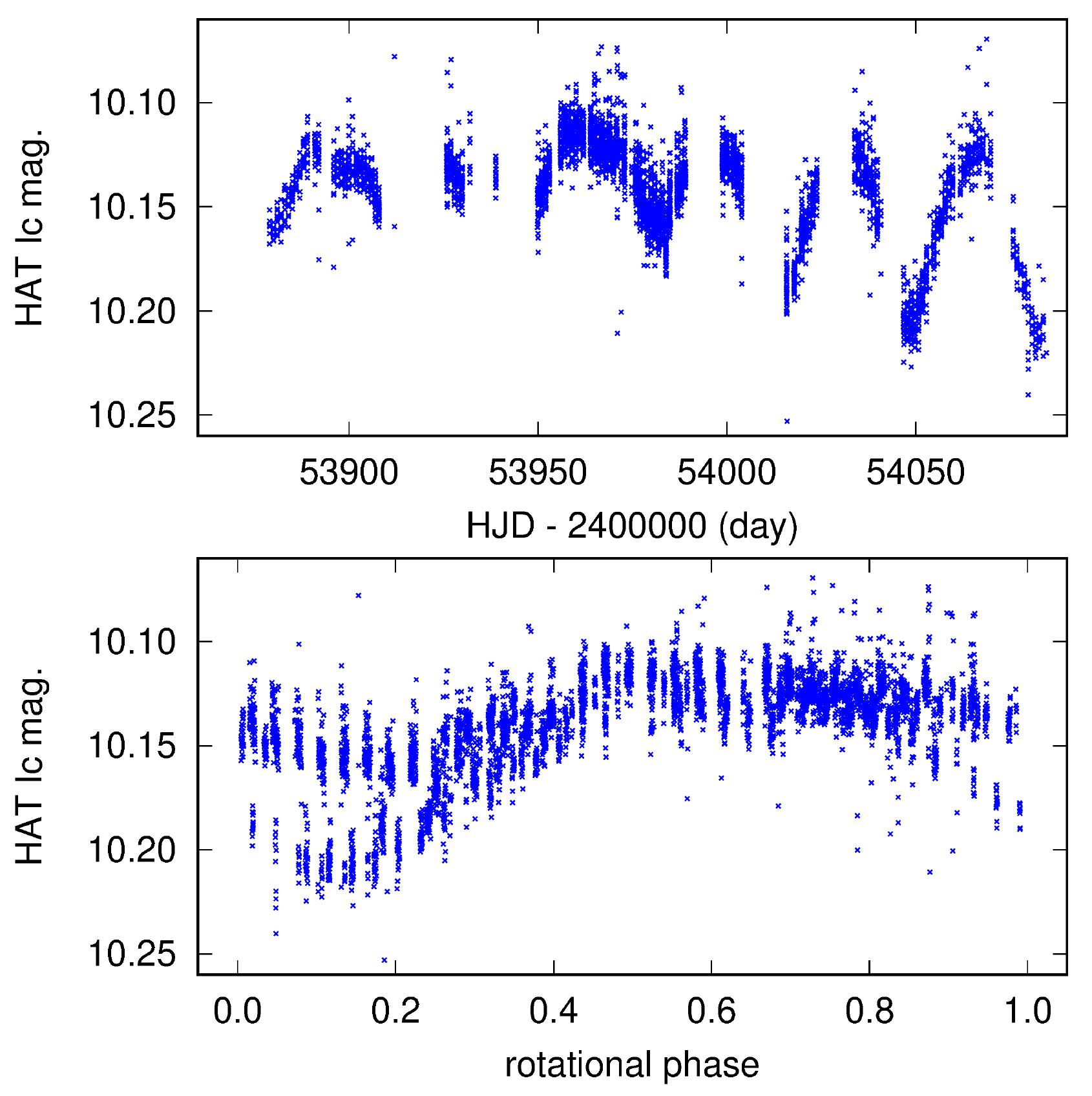}
   \caption{HATNet photometry of KIC\,2852961 collected in one observing season between May and December, 2006. Top panel shows the observations in $I$ band vs. J.D. (2,400,000.0+), while in the bottom panel the folded light curve is plotted with using $P_{\rm phot}=34.27$\,d period.}
              \label{fig1}
    \end{figure}

\begin{figure}
   \centering
   \includegraphics[width=\columnwidth]{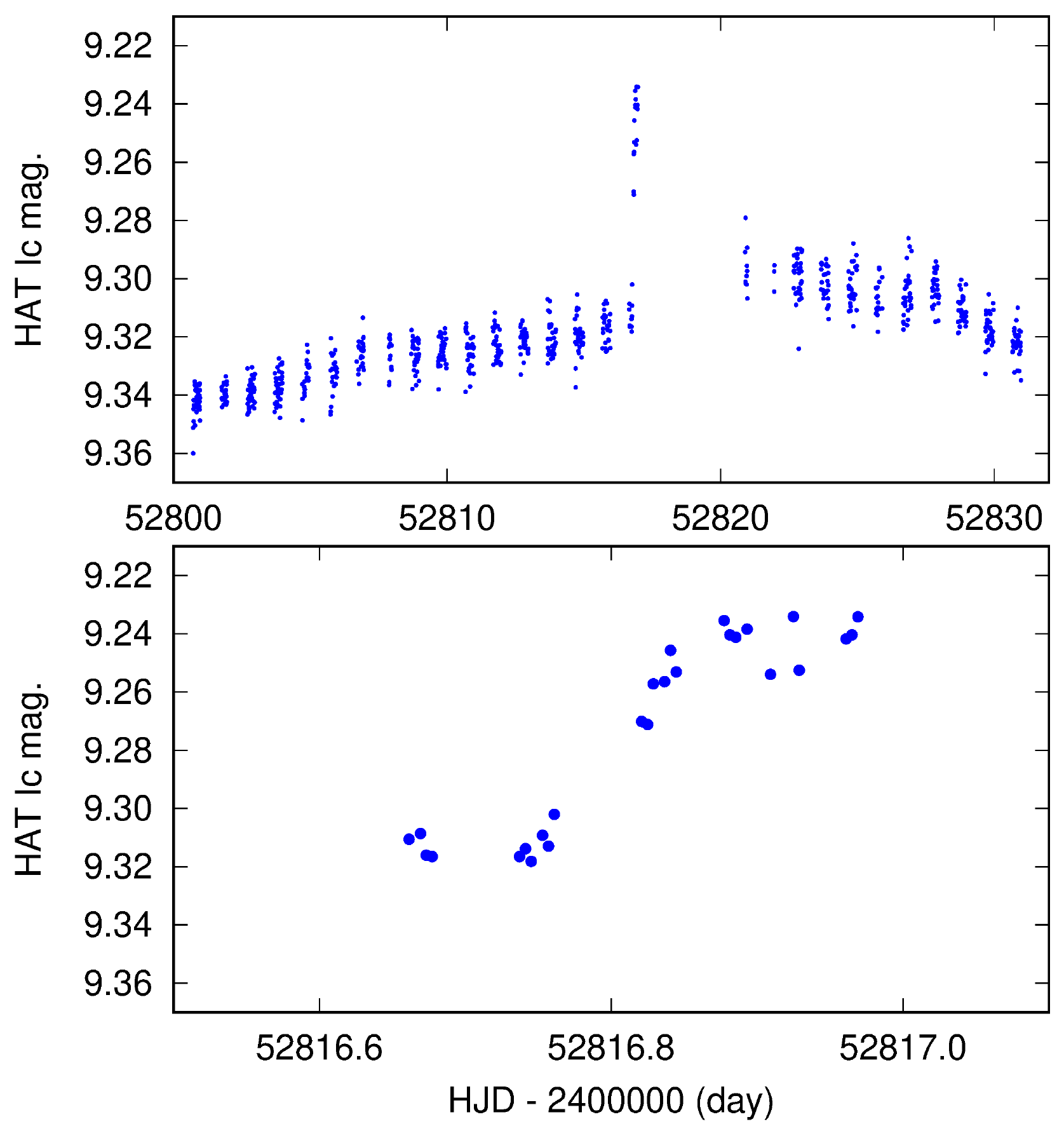}
   \caption{A large flare of KIC\,2852961 in 26 June, 2003, observed by HATNet. The slowly changing base light curve shown between 10 June and 10 July covers almost a whole rotation period. Apart from the flare, the volumes of the nightly changes indicate the overall scatter of the photometric measurements.}
              \label{fig2}
    \end{figure}

In our study we analyse the flaring activity of a target from the Kepler Input Catalogue under entry-name KIC\,2852961 (2MASS J19261136+3803107, TIC\,137220334)
using \emph{Kepler} and \emph{TESS} observations. The star was listed in the ASAS catalogue of variable stars in the Kepler field of view \citep{2009AcA....59...33P} with 35.58\,d rotational period, derived using 83 and 100 data points obtained in $V$ and $I$ colours, respectively, between May 28, 2006 and January 16, 2008. Figure~\ref{fig1} shows archival photometry from the Hungarian-made Automated Telescope Network \citep[HATNet,][]{2004PASP..116..266B} with 4475 datapoints in $I_C$ colour collected during the observing season in 2006 (205 days). The folded light curve underneath, assuming $P_{\rm phot}=34.27$\,d period, supports that the photometric period of $\approx$35 days is indeed due to rotation. Over and above, the shape of the slowly changing light curve is typical of spotted stars. From a little earlier period, in the 2003 HATNet dataset we found a large flare with an amplitude of about 0.08\,mag in the infrared, i.e., well above the noise limit of $\approx$0.01\,mag, unfortunately with sparse coverage of the decay phase; see Fig.~\ref{fig2}. Note that observing such an event from the ground is just the matter of blind luck.

%
According to the NASA Exoplanet Archive\footnote{\url{https://exoplanetarchive.ipac.caltech.edu}}
the low surface gravity of $\log g=2.919$, a surface temperature of 4722\,K and a radius of 5.5\,$R_{\odot}$ together with the $\approx$35\,d rotation period are all consistent with the preconception that KIC\,2852961 is a late G-early K giant. The \emph{Kepler} time series of our target were examined first by an automated Fourier-decomposition in \citet{2011A&A...529A..89D}, who found signatures of rotational modulation with other non-classified signs of variability, however, the presumed eclipsing binary nature was not confirmed. In turn, based on new high-resolution spectroscopic data  KIC\,2852961 has recently been categorized as a single-lined spectroscopic binary (SB1), but without more details \citep{2020arXiv200413792G}.
Our star is included in \citet[][see Table~2. in that paper]{2015MNRAS.447.2714B} as a rotational variable, 
but with a spurious period as the short-cadence \emph{Kepler} data did not cover the entire rotation. In the short-cadence data \citet{2015MNRAS.450..956B} searched for quasi-periodic pulsations induced by flares and found, that KIC\,2852961 indeed showed distinct \say{bumps} in the flare decay branches, which might be flare loop oscillations and unlikely the signs of induced global acoustic oscillation.

In this study
we use all the available \emph{Kepler} and \emph{TESS} light curves of KIC\,2852961 in order to search for stellar flares and study their occurrence rate, which is the only study of its kind for a flaring giant star so far. The paper is organized as follows. In  Sect.~\ref{radvel_obs} our new spectroscopic observations are presented and analysed. In Sect.~\ref{astrop} we revise the stellar parameters of KIC\,2852961. In Sect.~\ref{obs} we give a summary of the \emph{Kepler} and \emph{TESS} observations, while in Sect.~\ref{method} the applied data processing methods are described. The results are presented in Sections~\ref{results}--\ref{tempdist} and discussed in Sect.~\ref{disc}. Finally, a short summary with conclusion is given in Sect.~\ref{summary}.

   \begin{figure}
   \centering
   \includegraphics[width=\columnwidth]{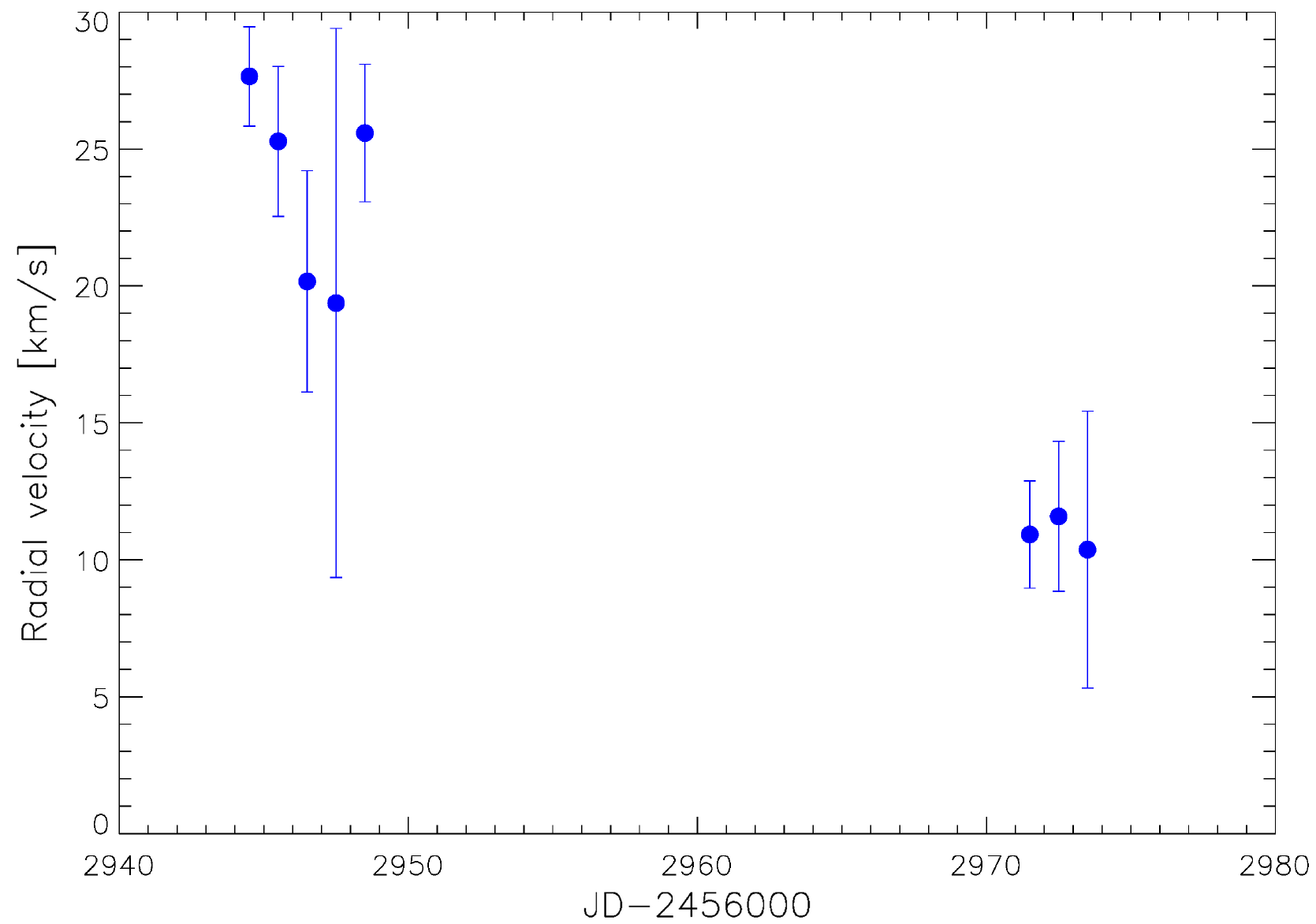}
   \caption{Radial velocity measurements of KIC\,2852961 taken by the 1-m RCC telescope at Piszkéstet\H{o} Mountain Station, Konkoly Observatory, Hungary, between 04 April and 03 May, 2020.}
              \label{fig_rv}
    \end{figure}
\section{Mid-resolution spectroscopic observations and data analysis}\label{radvel_obs}

New mid-resolution spectroscopic observations were taken between 04--08 April and 01--03 May, 2020 by the 1-m RCC telescope of Konkoly Observatory, located at Piszkéstet\H{o} Mountain Station, Hungary, equipped with a R=21000 \'echelle spectrograph. Altogether 27 spectra were collected with 2-5 exposures per day, depending on weather conditions. Circadian exposures were combined to get eight combined spectra with total exposures between 1-3 hours, yielding signal-to-noise ratios larger than 40 for each. The spectra were reduced using standard IRAF\footnote{\url{https://iraf.net}} \'echelle reduction tasks. Wavelength calibration was done using ThAr calibration lamps. The observational log is given in Table~\ref{tab_rv}.

 Radial velocity measurement were done in respect to the radial velocity standard HD\,159222 \citep{2018A&A...616A...7S}. The temporal variation of the radial velocity of KIC\,2852961 with the estimated error bars are given in the last two columns of Table~\ref{tab_rv} and plotted in Fig.~\ref{fig_rv}. Despite the different weather conditions which are reflected by the size of the error bars, the plot clearly shows that the radial velocity significantly decreased from $\approx$25\,kms$^{-1}$ in early April to $\approx$11\,kms$^{-1}$ in the beginning of May, i.e., during about 25 days. This change could indeed be the sign of orbital motion, in line with the SB1 nature suggested by \citet{2020arXiv200413792G}, but further observations are necessary to cover the full orbit.

Three of the best quality combined spectra were used to estimate the rotational broadening of the spectral lines with spectral synthesis. We apply the spectral synthesis code SME (Spectroscopy Made Easy, \citealt{piskunov_sme}). Atomic line data were taken from VALD database \citep{kupka_vald} and MARCS atmospheric models \citep{gustafsson_marcs} were used.
Keeping $T_{\rm eff}$, $\log g$ and [Fe/H] as free parameters the fits yielded $T_{\rm eff}$=4810$\pm$60\,K, 
$\log g$=2.49$\pm$0.10\,[cgs] and [Fe/H]=$-$0.25$\pm$0.10 with $v\sin{i}$=17.0$\pm$1.2\,kms$^{-1}$.
When $T_{\rm eff}$, $\log g$ and [Fe/H] are kept constant as 4722\,K, 2.43 and $-$0.08, respectively (see Table \ref{tab1} in Sect. \ref{astrop}), the spectral fits yielded $v\sin{i}\approx$18.0\,kms$^{-1}$ but with larger errorbars.
Herewith, therefore, we accept $v\sin{i}$=17.5\,kms$^{-1}$ but drawing attention to the fact that this result is preliminary due to the relatively low signal-to-noise ratio of the spectra and the smearing effect from the medium resolution.

\begin{table*}
   \centering
\caption{Observing log of the spectroscopic data taken by the Hungarian 1-m RCC telescope. Listed are the initial Julian Dates and dates of the observations, the exposure times and the numbers of subsequent exposures with the derived radial velocities and the corresponding errors .}
\label{tab_rv}
\begin{tabular}{c c c c r r}
\hline\noalign{\smallskip}
JD & Date & Exposure & Number of & $RV$ & $\sigma_{RV}$\\
start & dd.mm.yyyy & time [s] & exposures & [kms$^{-1}$] & [kms$^{-1}$] \\
\hline\hline\noalign{\smallskip}
2458944.4672 & 04.04.2020 & 3600 & 3 & 27.65 & 1.82 \\
2458945.4630 & 05.04.2020 & 3600 & 3 & 25.28 & 2.74 \\
2458946.5627 & 06.04.2020 & 1800 & 3 & 20.16 & 4.04 \\
2458947.4635 & 07.04.2020 & 3600 & 3 & 19.38 & 10.03 \\
2458948.5754 & 08.04.2020 & 1800 & 2 & 25.58 & 2.51 \\
2458971.3811 & 01.05.2020 & 1800 & 5 & 10.92 & 1.96 \\
2458972.4231 & 02.05.2020 & 3600 & 3 & 11.59 & 2.74 \\
2458973.4620 & 03.05.2020 & 3600 & 3 & 10.37 & 5.06 \\
\hline
\end{tabular}
\end{table*}

\section{Revised stellar parameters of KIC\,2852961}\label{astrop}

The \emph{Gaia} DR-2 parallax of 1.2845$\pm$0.0259\,mas \citep{2018A&A...616A...1G} with taking into account a mean offset of --53.6\,$\mu$as \citep[cf. Fig. 1 in][and their references]{2019ApJ...878..136Z} yields a distance of 813$\pm$17\,pc for our target. According to the ASAS light curve \citep[][]{2009AcA....59...33P} the brightest ever observed visual magnitude can be estimated as $V_{\rm max}\approx10\fm30$. This, together with the above mentioned distance and an interstellar extinction of $A_V$=$0\fm264$ from 2MASS \citep[][]{2006AJ....131.1163S} gives an absolute visual magnitude of $M_{V}$=$0\fm486\pm0\fm046$. The effective temperature $T_{\rm eff}$=$4722^{+77}_{-56}$\,K from the Kepler Input Catalogue \citep[][]{2009yCat.5133....0K} (which is practically the same as $4739^{+101}_{-93}$\,K from \emph{Gaia} DR-2) is consistent with a bolometric correction of $BC$=$-0\fm454^{+47}_{-37}$ \citep{1996ApJ...469..355F}. This yields a bolometric magnitude of $M_{\rm bol}$=$0\fm032\pm0\fm090$, convertible to $L=76.5^{+6.0}_{-6.3}\,L_{\odot}$. Taking this luminosity value the Stefan-Boltzmann law would give $R$=13.1$\pm$0.9\,$R_{\odot}$. This new radius is more than two times bigger than the one listed in the NASA Exoplanet Archive, however, it is derived in a trustworthy way and it is in a better agreement with the radius of 10.64\,$R_{\odot}$ given in \emph{Gaia} DR-2 \citep{2018A&A...616A...1G}. The surface temperature and the derived luminosity are used to plot our target on the Hertzsprung--Russell (H-R) diagram in Fig.~\ref{hrd}. Stellar evolution tracks are taken from Padova and Trieste Stellar Evolution Code ({\it PARSEC}, \citealt[][]{2012MNRAS.427..127B}) for Z=0.01 ([M/H]=$-$0.175). From Fig.~\ref{hrd} we estimate a stellar mass of 1.7$\pm 0.2$\,$M_{\odot}$ for KIC\,2852961 with an age of $\approx$1.7\,Gyr, i.e. around the red giant bump. We note however, that due to the uncertainty in metallicity the true errors should be $\approx$50\% larger compared with those estimated from purely $T_{\rm eff}$ and luminosity. The above mentioned mass and radius would yield a surface gravity of $\log g$=2.43$\pm$0.14. Albeit this is smaller by $\approx$20\% than the value of 2.919$\pm$0.145 given in the NASA Exoplanet Archive, it is more consistent with the revised astrophysical properties. Finally, taking $v\sin{i}$ of 17.5\,kms$^{-1}$ obtained from spectral synthesis (see Sect. \ref{radvel_obs}) with the maximum equatorial velocity of 18.6\,kms$^{-1}$ calculated from the photometric period and the estimated radius would yield $\approx$70$^{\circ}$ inclination. The most important stellar parameters of KIC\,2852961 are listed in Table\ref{tab1}.

   \begin{figure*}[t!!!!]
   \centering
   \includegraphics[width=2\columnwidth]{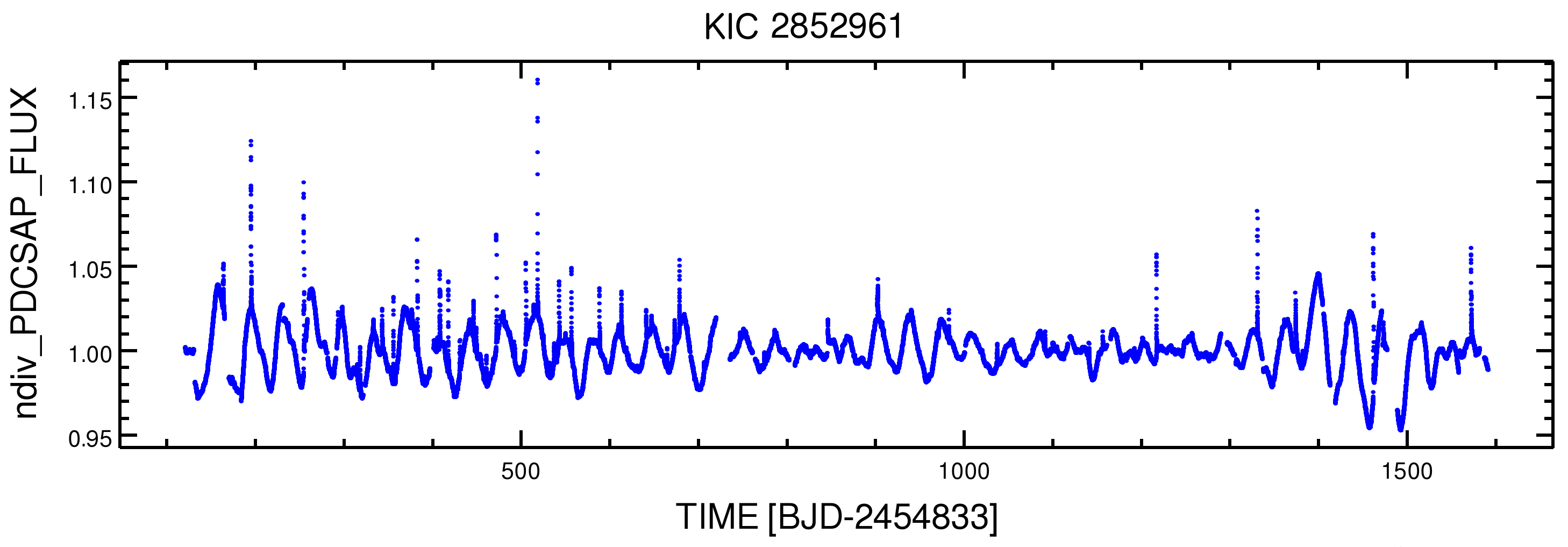}
   \caption{Long-cadence \emph{Kepler} data of KIC\,2852961. Note the rotational variability which is changing from one rotational cycle to the next, typical to stars with constantly renewing spotted surface. Several flare events are also present including really big ones.}
              \label{fig3}
    \end{figure*}

   \begin{figure}[thb]
   \centering
   \includegraphics[width=\columnwidth]{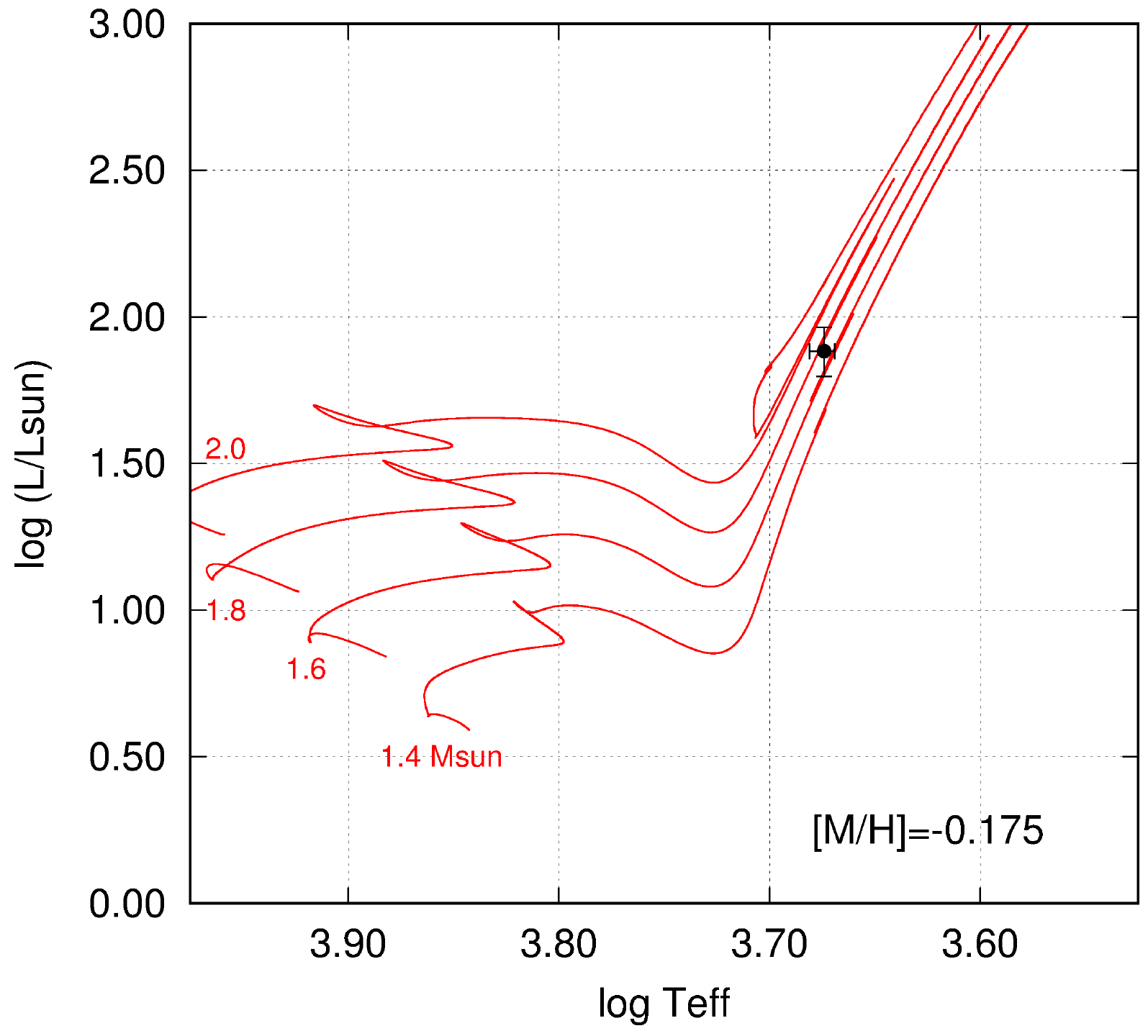}
   \vspace{-0.5cm}
   \caption{Position of KIC\,2852961 in the H-R diagram (dot). Stellar evolutionary tracks shown between 1.4 and 2.0 solar masses by 0.2 steps are taken from {\it PARSEC} \citep{2012MNRAS.427..127B} adopting [M/H]=$-$0.175. From the location the expected mass is around 1.7\,$M_{\odot}$.}
              \label{hrd}
    \end{figure}

 \begin{center}
 \begin{table}[thb]
\caption{Revised astrophysical data of KIC\,2852961} \label{tab1}
 \begin{tabular}{ll}
  \hline\noalign{\smallskip}
  Parameter               &  Value \\
  \hline\hline\noalign{\smallskip}
  Spectral type            & $\sim$G9-K0 III  \\
  Distance [pc]\tablefootmark{a} &  $813\pm{17}$\\
  $V_{\rm max}$ [mag]\tablefootmark{b} & $\approx$$10\fm30$ \\
  $M_{\rm bol}$     [mag]    & $0\fm032\pm0\fm090$ \\
  Luminosity [${L_{\odot}}$]         & $76.5^{+6.0}_{-6.3}$  \\
  $T_{\rm eff}$ [K]\tablefootmark{c} &           $4722^{+77}_{-56}$    \\
  Radius [$R_{\odot}$] &      $13.1\pm 0.9$   \\
  Mass [$M_{\odot}$]   & $1.7\pm0.3$   \\
  $\log g$ [cgs]  &   $ 2.43\pm0.14$ \\
  Metallicity [Fe/H]\tablefootmark{c} &  $-0.08^{+0.15}_{-0.1}$ \\
  $v\sin{i}$ [kms$^{-1}$]  &   $\approx$17.5 \\
  Inclination [$^{\circ}$]  & $70\pm10$ \\
  Photometric  period [d]\tablefootmark{d}   &  $\approx$$35.5$ \\
  \hline\noalign{\smallskip}
\end{tabular}\\
\tablefoot{
\tablefoottext{a}{Taken from \emph{Gaia} DR-2.}
\tablefoottext{b}{Taken from ASAS Archive.}
\tablefoottext{c}{Taken from NASA Exoplanet Archive.}
\tablefoottext{d}{Computed by the Periodogram Tool of the NASA Exoplanet Archive.}
}
\end{table}
\end{center}


\section{Observations from \emph{Kepler} and \emph{TESS}}\label{obs}

\begin{figure}[th]
   \centering
   \includegraphics[width=\columnwidth]{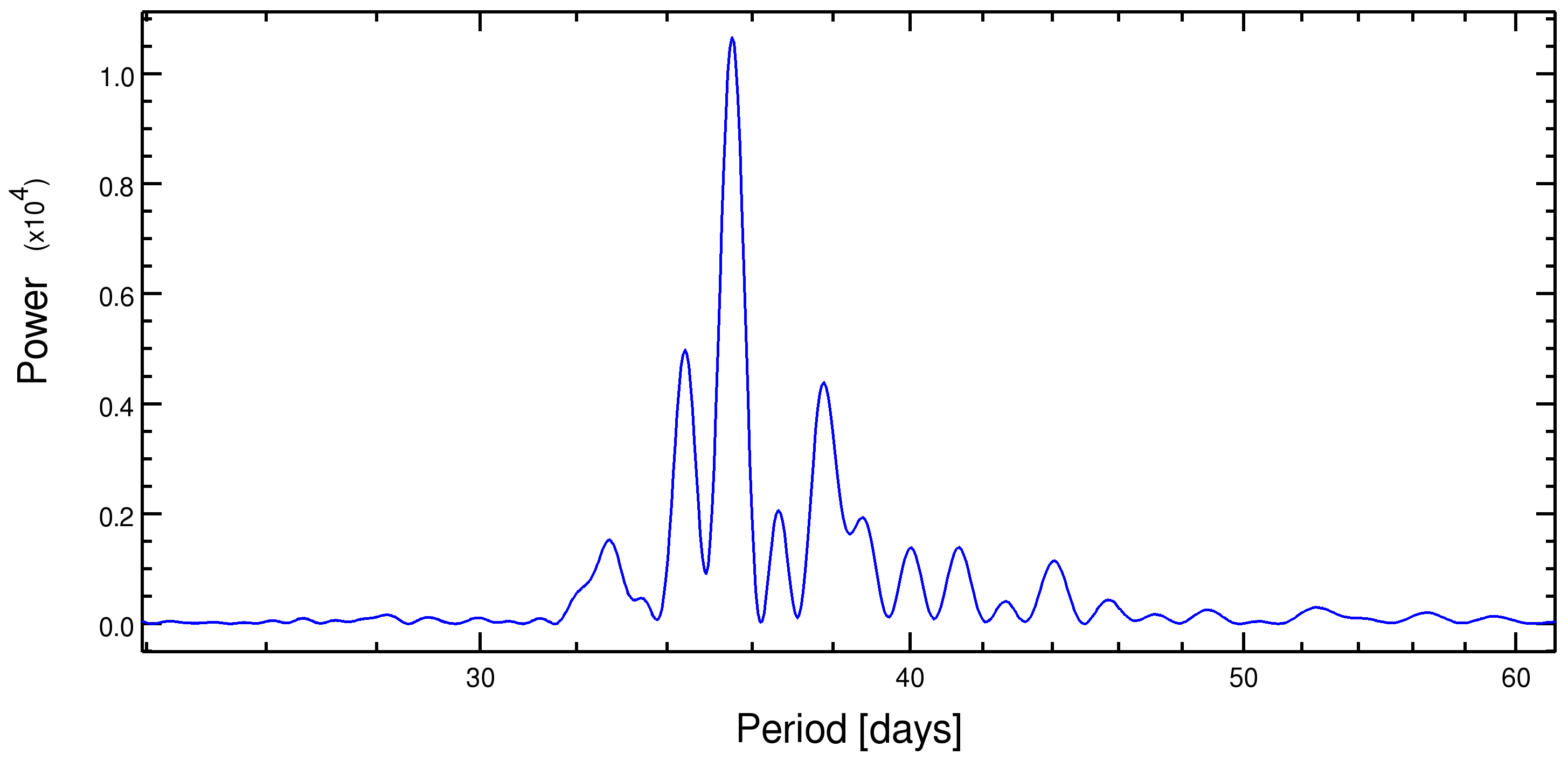}
   
   \vspace{0.1cm}
   \includegraphics[width=\columnwidth]{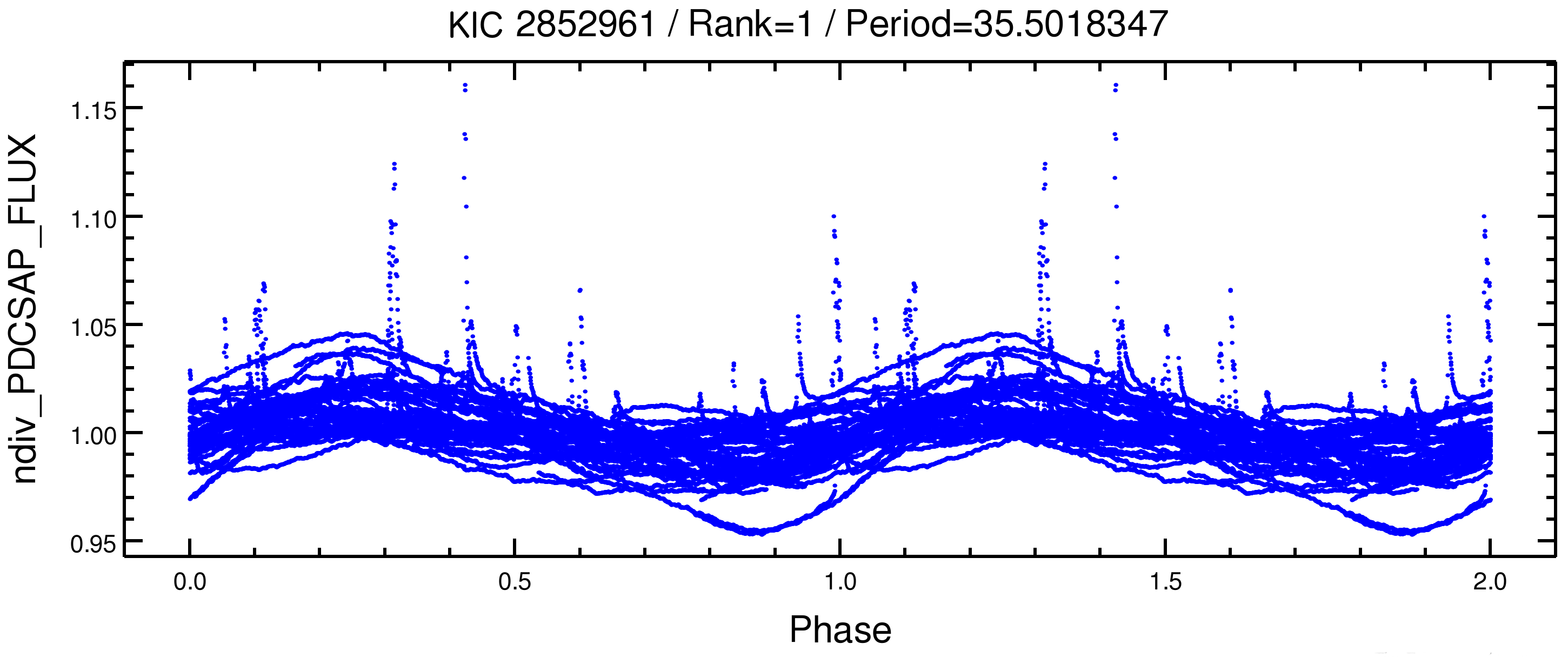}
   
   \vspace{-0.1cm}
   \caption{Periodogram (top) and folded light curve (bottom) of KIC\,2852961 using data shown in Fig.\ref{fig3}. The highest peak of the Lomb-Scargle periodogram indicates a rotation period of $\approx$35.5\,d. Plots are created with the Periodogram Tool of the NASA Exoplanet Archive.}
              \label{fig4}%
    \end{figure}

The full \emph{Kepler} dataset available for KIC\,2852961 in NASA Exoplanet Archive was collected between BJD\,2454953.5 and BJD\,2456424.0. A total of 19 \emph{Kepler} light curves were observed, of which 18 long cadence time series from quarters Q0-Q17 with a time resolution (effective integration time) of 30 minutes, plus one short cadence light curve in Q4 with one minute sampling.
The normalized long cadence data are plotted in Fig.~\ref{fig3}, indicating the rotational modulation due to spots together with strong flare activity. This time series was taken in order to search for the best-fit photometric period. For the computations we used the Lomb-Scargle algorithm \citep{1976Ap&SS..39..447L,1982ApJ...263..835S} in the Periodogram Tool of the NASA Exoplanet Archive. According to the resulting periodogram in Fig.~\ref{fig4} the strongest peak is located at $P_{\rm phot}\approx35.5$\,d, i.e., in a pretty good agreement with other period determinations using ASAS and HATNet datasets (see Sect.~\ref{intro}). We notice that the bunch of peaks in the periodogram in Fig.~\ref{fig4}, as well as the different main periods found for other datasets (ASAS, HATNet) observed at different times might be the indication of surface differential rotation.

Additionally, we use \emph{TESS} Sector 14 data taken between July 18 and August 15, 2019 (BJD\,2454953.5--2456424.0). The 30-minute cadence data were extracted using the \texttt{eleanor} pipeline\footnote{\url{https://adina.feinste.in/eleanor/}}, an open-source tool to produce light curves for objects in the \emph{TESS} Full-Frame Images \citep{2019PASP..131i4502F}.

   \begin{figure*}[thb]
   \centering
   \includegraphics[width=2\columnwidth]{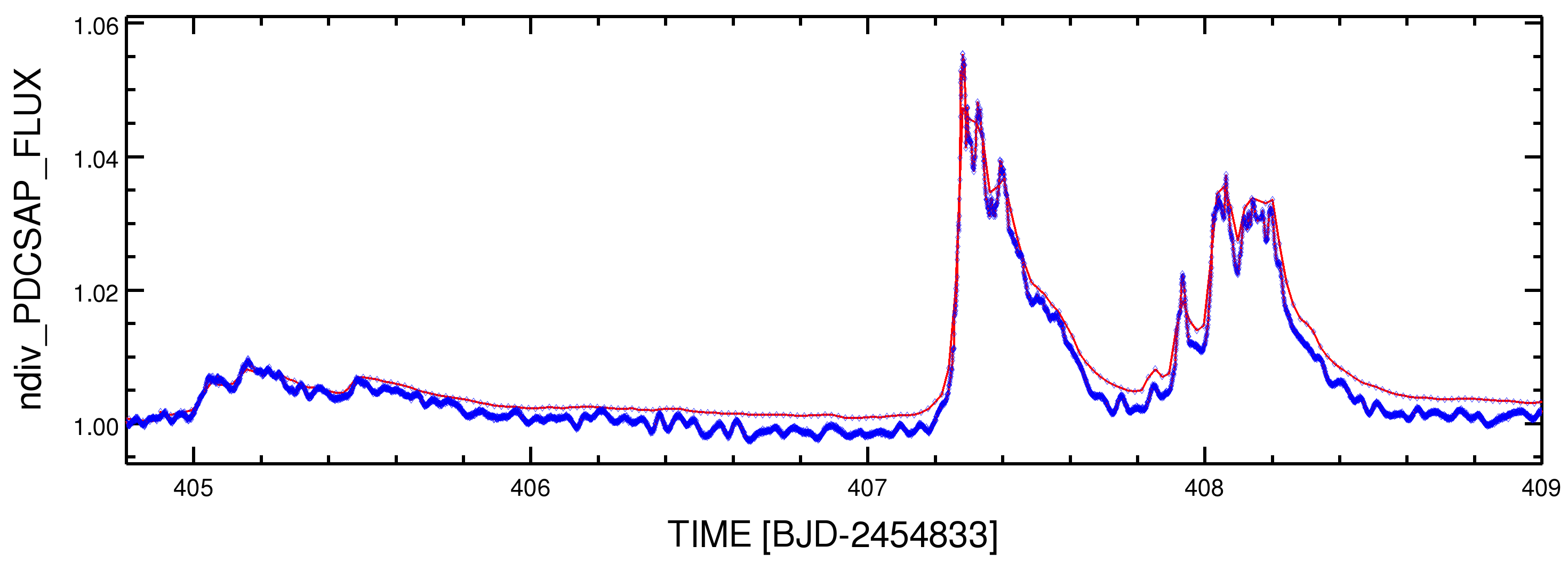}
   \caption{Three flares observed on KIC\,2852961 with short and long cadence in \emph{Kepler} Q4. Short-cadence data are drawn with thick blue line, while long-cadence data are overplotted with thin line in red. }
              \label{fig_sclc}
    \end{figure*}

\section{Detecting flares in photometric data from space}\label{method}

We search for flares using an automated technique accompanied by visual inspection. For this, we apply an updated version of the flare detection pipeline from \citet{2020AJ....159...60G}. We start from the detrended \emph{Kepler} and \emph{TESS} light curves and compute a Lomb-Scargle periodagram to identify (semi-)periodic modulation caused by stellar variability or rotation. 
The strongest periodic signal is removed using a spline with knots spaced at one tenth of the detected period. 
At the same time, we search for outliers by sigma clipping the data residuals, identifying and masking all outliers that are more than 3-$\sigma$ away.
We repeat this entire process two more times or until no new periods are found, collecting a list of all outliers.
These outliers are considered to be flare-like events if they contain a series of at least three consequent 3-$\sigma$ outlier points. 

These candidates are then re-examined by eyeball to decide if they have a flare-like profile and can be declared a flare. To this end, the classical flare profile with a rapid rise followed by exponential decay branch helped in the verification. 
However, various doubtful cases were found at the low energy end, where quasi-oscillations, scatter and instrumental glitches either hampered the detection or caused false positives.
At the end we confirmed 59 flare events in the \emph{Kepler} time series (three of them were observed simultaneously in both short and long cadence data), while one single event was found in the much shorter \emph{TESS} light curve.

\section{Calculation of the flare energies}\label{results}

For each confirmed flare events we calculate flare energies in the following way. Let ${I_{0}}$ and ${I_{0+f}}$ be the intensity values (either bolometric or in a given bandpass) of the stellar surface at quiescent and flaring state, respectively. The flare energy relative to the star can be written as
\begin{equation} 
\varepsilon_f=\int_{t_1}^{t_2} (\frac{I_{0+f}(t)}{I_0} - 1) dt,
\end{equation}
where $t_1$ and $t_2$ are the start an the end points of the given flare event. The quiescent stellar flux is estimated using black body approximation, where Planck's law gives the spectral radiance:
\begin{equation}
{B}(\lambda,T)=\frac{2 h c^2}{\lambda^5}\frac{1}{{\rm exp}(\frac{hc}{\lambda k T}-1)}.
\end{equation}
By integration over all solid angles of a hemisphere the spectral exitance is $\pi {B}(\lambda,T)$, therefore the quiescent stellar luminosity through the \emph{Kepler} filter would read
\begin{equation}
L_{\star\rm Kep}= A_\star\int_{\lambda_1}^{\lambda_2} \pi B(\lambda,T) \mathcal{K}(\lambda) d\lambda,
\end{equation}
where $A_\star$ is the stellar surface while $\mathcal{K}(\lambda)$ is the \emph{Kepler} response function given between $\lambda_1$ and $\lambda_2$. At this point the total integrated
flare energy through the \emph{Kepler} filter is obtained by multiplying the relative flare energy by the quiescent stellar luminosity:
\begin{equation}
E_f=\varepsilon_f L_{\star\rm Kep}.
\end{equation}

Taking $R$ and $T_{\rm eff}$ from Table~\ref{tab1} the quiescent stellar luminosity through the \emph{Kepler} filter is $L_{\star\rm Kep}=7.2\times10^{34}$\,$\rm{erg\,s^{-1}}$. (In comparison, the total bolometric luminosity from the Stefan-Boltzmann law is $L_\star=2.85\times10^{35}$\,$\rm{erg\,s^{-1}}$.) The calculated $E_f$ flare energies with the corresponding $T_{\rm peak}$ peak times and $\Delta t$ durations are listed in Table~\ref{tab2} together with the equivalent flare duration ($t_{\rm eq}$) values. Note, that this latter is identical with $\varepsilon_f$ from Eq.~1, since it is basically the time interval in which the star would radiate as much energy as the given flare itself, that is
\begin{equation}
t_{\rm eq}\equiv\varepsilon_f =\frac{E_f}{L_{\star\rm Kep}}.
\end{equation}\label{eqen}

In Quarter 4 \emph{Kepler} season, where both low and short cadence data are available, the calculations are performed for three confirmed flare events detected simultaneously in both datasets; for the three flares see Fig.~\ref{fig_sclc}. The derived corresponding flare energies (see the 13--15th rows in Table~\ref{tab2}) are quite close to each other, differing by 1--6\% only. Therefore we estimate that the uncertainty of our flare energy calculations is within $\approx$10\%. On the other hand, flare energies can also be calculated by assuming black body continuum emission at around $T_{\rm BB}$=8000$-$10000\,K \citep[see e.g.][and the references therein]{2011A&A...530A..84K}. Using the method in \citet{2013ApJS..209....5S} we recalculated some of our flare energies with assuming $T_{\rm BB}$=9000\,K, which yielded $\approx$8\% difference compared to the values in Table~\ref{tab2} (i.e., still within the 10\% uncertainty). Best match was obtained when $T_{\rm BB}$$\approx$8300\,K was used.

The calculations above are repeated for the \emph{TESS} flare but using the \emph{TESS} transmission function instead; for the results see Table~\ref{tab3}. We note that since $t_{{\rm eq},TESS}$ and $E_{f,TESS}$ values in Table~\ref{tab3} are calculated using a different filter function, therefore they cannot be compared directly to the respective values of the \emph{Kepler} flares in Table~\ref{tab2}.

\begin{center}
\begin{table}
\caption{Peak times (BJD--2454833), flare durations, equivalent durations and flare energies in the \emph{Kepler} bandpass. $E_{f,{\rm SC}}$ values are from Q4 short-cadence. Complex flares have been marked with $\ast$ symbols.}
\label{tab2}
\vspace{-0.275cm}
\begin{tabular}{r r r r r}
\hline\hline\noalign{\smallskip}
$T_{\rm peak}$  & $\Delta t$\,[h]  & $t_{\rm eq}$\,[sec]  & $E_f$\,[erg] &  $E_{f,{\rm SC}}$\,[erg]\\
\hline\noalign{\smallskip}
163.329    &  33.60  &  936.879  & 6.746e+37  & \\
$\ast$ 194.634  &  36.00  &  3157.145  &  2.273e+38 &  \\
$\ast$ 254.116    &  37.92  &  3330.501  &  2.398e+38  & \\
$\ast$ 293.142    &  6.24  &  56.023  &  4.034e+36  & \\
$\ast$ 314.923    &  26.40  &  216.666  &  1.560e+37  &  \\
$\ast$ 317.620    &  48.00  &  960.905  &  6.919e+37   & \\
$\ast$ 332.659    &  31.20  &  336.170  &  2.421e+37   & \\
$\ast$ 342.548    &  39.60  &  847.920  &  6.106e+37  & \\
344.367    &  9.60  &  18.327  &  1.320e+36  & \\
355.155    &  24.00  &  1023.951  &  7.373e+37 &  \\
$\ast$ 376.242    &  18.72 &  222.661  &  1.603e+37  & \\
$\ast$ 382.290    &  29.76   &  1261.709  &  9.085e+37  & \\
$\ast$ 405.156    &  27.12  &  328.035  &  2.362e+37  & 2.500e+37 \\
407.281    &  15.84  &  875.386  &  6.304e+37  & 6.354e+37 \\
$\ast$ 408.057    &  27.36  &  822.838  &  5.925e+37  & 5.562e+37 \\
$\ast$ 417.212    &  19.20  &  1060.302  &  7.635e+37 &   \\
430.330    &  22.08  &  367.232  &  2.644e+37   & \\
$\ast$ 445.779    &  31.20  &  690.827  &  4.975e+37  & \\
$\ast$ 448.946    &  24.00  &  393.929  &  2.837e+37   & \\
460.144    & 30.00   &  459.364  &  3.308e+37   & \\
471.506    &  21.60  &  1146.975  &  8.259e+37   &  \\
$\ast$ 479.230    &  14.88  &  80.285  &  5.781e+36   & \\
481.968    &  18.72  &  60.677  &  4.369e+36   & \\
$\ast$ 491.634    &  7.92  &  66.901  &  4.817e+36   & \\
504.876    &  20.40  &  743.118  &  5.351e+37  &  \\
518.015    &  16.80  &  1633.223  &  1.176e+38  &  \\
531.154    &  6.72  &  35.933  &  2.588e+36   & \\
536.794    & 23.28   &  167.759  &  1.208e+37   & \\
$\ast$ 542.495    &  22.80  &  736.110  &  5.301e+37   & \\
549.627    &  16.08  &  75.842  &  5.461e+36   & \\
556.288    &  28.80  &  1259.140  &  9.067e+37  &  \\
588.001    &  25.20  &  699.830  &  5.039e+37   & \\
$\ast$ 609.661    &  8.88  &  42.707  &  3.075e+36   & \\
$\ast$ 612.726    &  32.88  &  773.672  &  5.571e+37  &  \\
$\ast$ 640.719    &  34.56  &  498.456  &  3.589e+37   & \\
646.869    & 13.20   &  82.132  &  5.914e+36   & \\
664.053    &  20.40  &  154.273  &  1.111e+37  &  \\
672.839    &  14.40  &  228.669  &  1.647e+37   & \\
678.192    &  34.80  &  914.100  &  6.582e+37   & \\
$\ast$ 774.105    &  21.60  &  269.828  &  1.943e+37   & \\
785.854    &  8.40  &  33.105  &  2.384e+36   & \\
787.489    &  7.44  &  16.188  &  1.166e+36   & \\
$\ast$ 902.126    &  28.80  &  866.239  &  6.238e+37   & \\
$\ast$ 902.923    &  5.28  &  10.871  &  7.828e+35   & \\
903.638    &  8.40  &  13.254  &  9.544e+35   & \\
908.890    &  6.96  &  29.748  &  2.142e+36   & \\
982.104    &  13.68  &  180.127  &  1.297e+37   & \\
$\ast$ 1036.700   &  31.20  &  354.456  &  2.552e+37   & \\
1090.908    &  12.48  &  182.304  &  1.313e+37   & \\
1140.459    &  11.28  &  144.011  &  1.037e+37   & \\
1156.172    & 6.96   &  79.607  &  5.732e+36   & \\
$\ast$ 1191.687    &  10.32  &  77.439  &  5.576e+36   & \\
$\ast$ 1216.617    &  26.40  &  1312.973  &  9.455e+37   & \\
1330.395    &  23.52  &  1412.387  &  1.017e+38   & \\
1381.805    &  14.16  &  163.963  &  1.181e+37   & \\
$\ast$ 1461.451    &  27.60  &  2501.156  &  1.801e+38   & \\
1472.914    &  22.32  &  321.986  &  2.319e+37   & \\
1566.071    &  5.28  &  7.629  &  5.493e+35  &  \\
1571.793    &  32.40  &  1564.891 & 1.127e+38 & \\
\hline\noalign{\smallskip} 
\end{tabular}\\
\end{table}
\end{center}
\begin{center}
\begin{table}
\caption{Peak time, duration, equivalent duration and integrated flare energy in the \emph{TESS} bandpass for the only flare detected in the \emph{TESS} Sector 14 data.}
\label{tab3}
\begin{tabular}{c c c c}
\hline\hline\noalign{\smallskip}
$T_{\rm peak}$  & $\Delta t$ & $t_{{\rm eq},TESS}$ & $E_{f,TESS}$ \\
 BJD--2458000 & [h]  & [sec]  & [erg] \\
\hline\noalign{\smallskip}
694.951  & 38.832  & 984.656  & 9.925e+37 \\
\hline\noalign{\smallskip} 
\end{tabular}\\
\end{table}
\end{center}

Among the flare events listed in Tables~\ref{tab2} and \ref{tab3} the shortest ones last approximately 5 hours, while the longest one reached a length of two days, which is extremely long. The respective integrated energy values range between $10^{35}$ and $10^{38}$ ergs, i.e. they span over three orders of magnitude. In our case few times $10^{35}$\,erg should be regarded as the detection limit due to data noise and other reasons (e.g., quasi-oscillations which slightly vary the overall brightness on a timescale of few hours). On the high end of the energy range some of the events show quite unusual structure. Such an event can be the result of multiple (regular or irregular) flare eruptions emerging at the same time, either in physical connection or independently by coincidence. However, other peculiarities should also be considered, such like the quasi-periodic pulsations (QPPs) in the decay phase of a flare  \citep{2015MNRAS.450..956B,2016MNRAS.459.3659P,2019ApJS..244...44B}.  The flares observed in short-cadence (see our Fig.\ref{fig_sclc}) were reported to show QPPs in \citet{2016MNRAS.459.3659P} and \citet{2015MNRAS.450..956B}.

   \begin{figure}[thb!!!!!]
   \centering
   \includegraphics[width=\columnwidth]{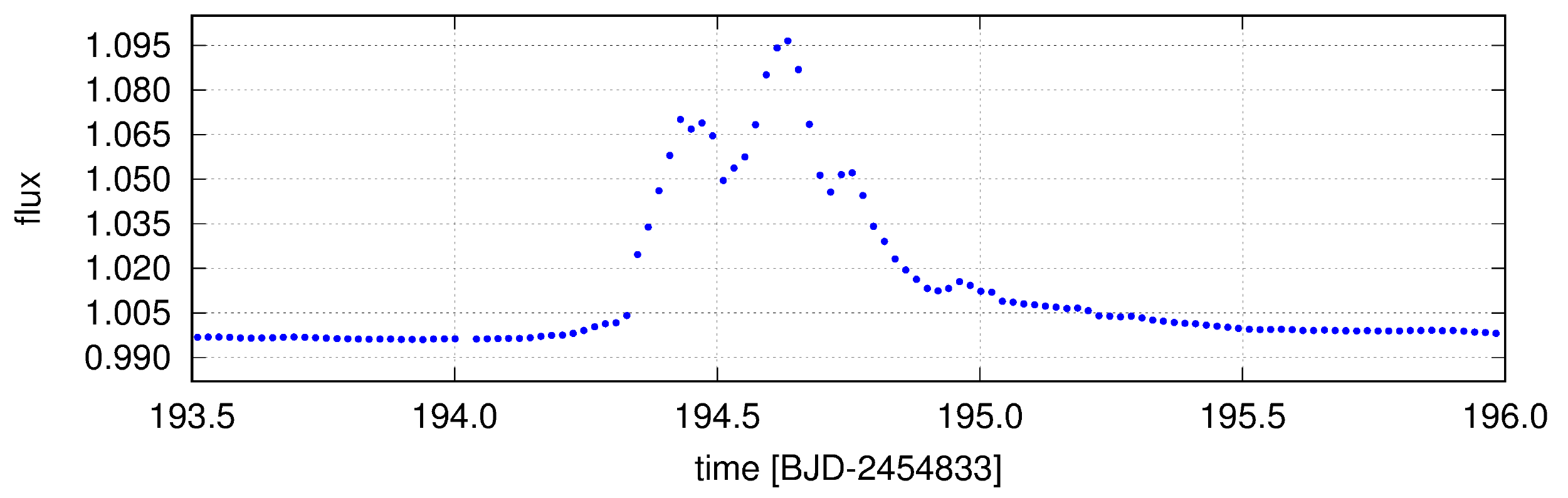}
   
   \vspace{0.4cm}
   \includegraphics[width=\columnwidth]{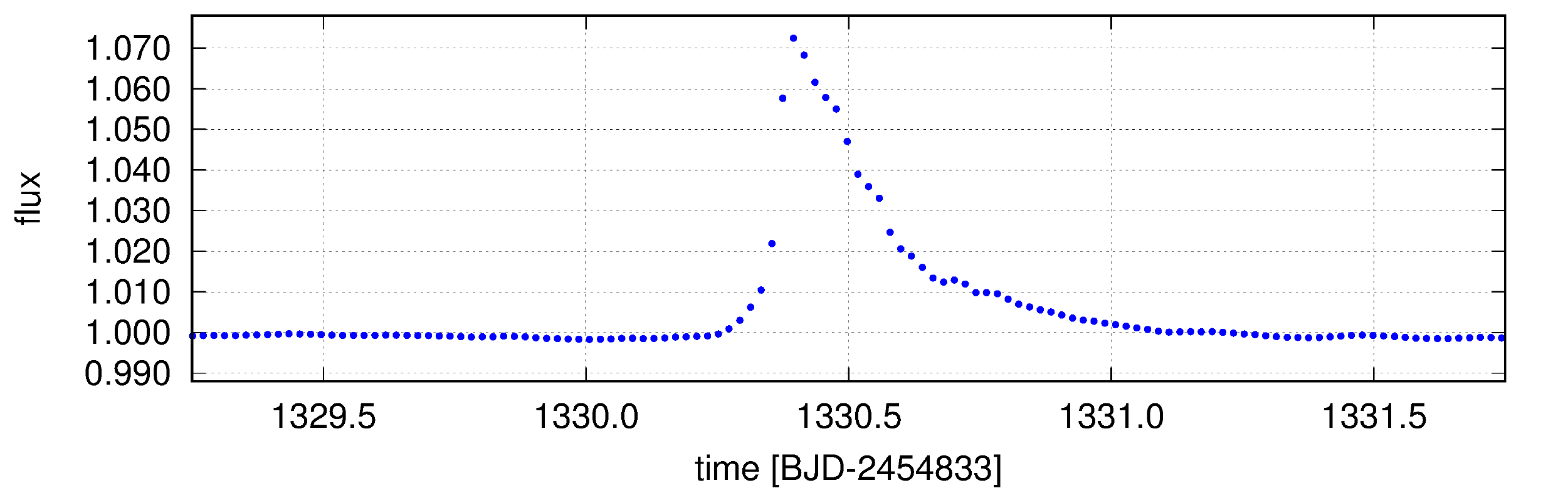}
   \caption{Two examples for flare detections on KIC\,2852961 by \emph{Kepler}. Top panel shows a (very likely multiple) flare event of $E_f=2.273\,10^{38}$\,erg energy (find at $T_{\rm peak}=194.634$\,d in Table~\ref{tab2}) with a complex, irregular shape, while the superflare of $E_f=1.017\,10^{38}$\,erg  shown in the bottom panel (find at $T_{\rm peak}=1330.395$\,d in Table~\ref{tab2}) has a regular light curve.}
              \label{twoflares}%
    \end{figure}
   \begin{figure}[ht!!!!]
   \centering
   \includegraphics[width=\columnwidth]{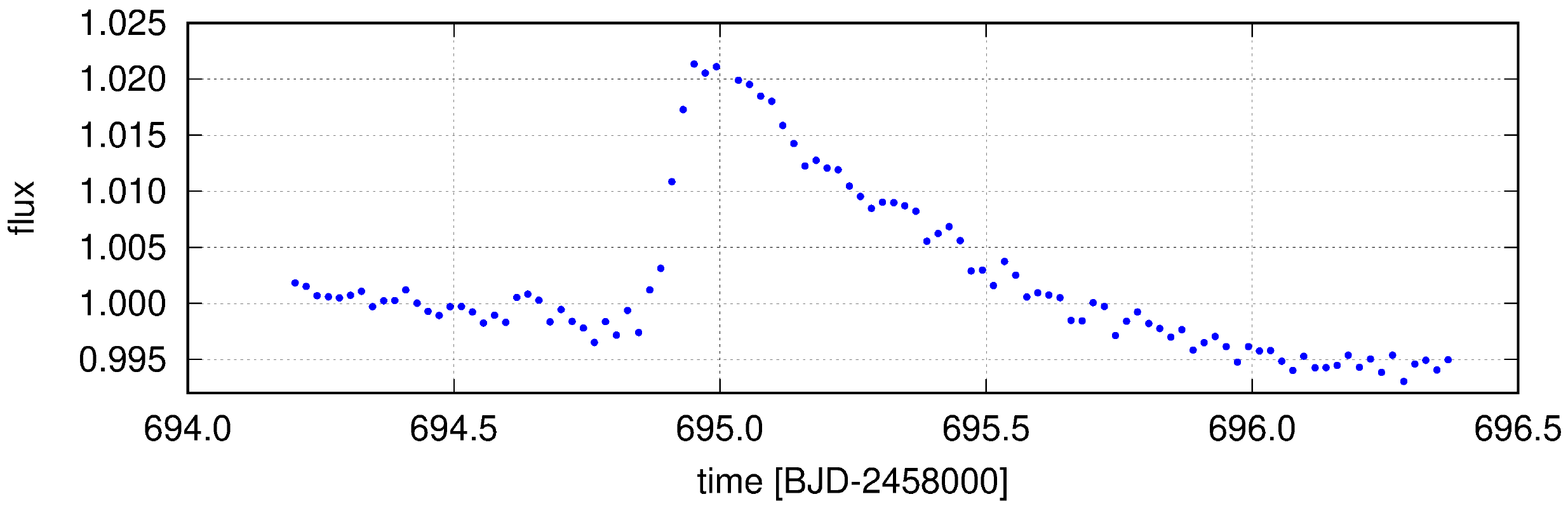}
   \caption{The only flare event on KIC\,2852961 found in our \emph{TESS} Sector 14 data, showing a fairly regular shape; cf. Table~\ref{tab3}.}
              \label{fig7_TESSflare}
    \end{figure}

From the list of flares in Table~\ref{tab2} we show two more examples in Fig.~\ref{twoflares}: one with a complex structure and another one with a regular shape. The decay phase of the complex flare very likely shows indications of QPPs \citep[cf.][]{2016MNRAS.459.3659P}, even in long-cadence sampling.  We note that complex events (marked with $\ast$ symbols in Table~\ref{tab2}) similar to the plotted one might be the superimposition of a couple of simultaneous flares as well. Nevertheless, in our analysis such a complex flare is regarded as one single event, because in most cases it is hardly possible to differentiate a real complex event from a series of individual flares overlapping each other in time. On the other hand, it is very likely, that such simultaneous flare events are physically connected by sharing the same active region and maybe triggering each other as well \citep{2008A&A...488L..29L}, i.e., releasing magnetic energy from the same resource, therefore it is reasonable to handle them as one event. As a third example, in Fig.~\ref{fig7_TESSflare} we show the only flare detected in the \emph{TESS} light curve (see Table~\ref{tab3}), having a regular shape and lasting almost forty hours. The emitted energy of $\approx$$10^{38}$\,erg in the \emph{TESS} bandpass (visible-near infrared spectrum) classifies it as one of the most powerful superflares of KIC\,2852961 (cf. Table~\ref{tab2}).

\section{The cumulative flare frequency distribution}\label{statistics}

In order to investigate the dependence of flare occurrence on flare energy we follow the method introduced by \citet{1972Ap&SS..19...75G}, who found that cumulative flare frequency distribution (hereafter FFD) for flare stars tended to follow a power law \citep[see also][]{1976ApJS...30...85L}. According to that, the $\Delta N(E)$ number of flares in the energy range $E+\Delta E$ per unit time (days) can be written as
\begin{equation}
\Delta N(E)\propto E^{-\alpha}\Delta E.
\end{equation}
Rewriting in a differential form and integrating between $E$ and $E_{\rm max}$ (i.e., the cutoff energy) one gets that the $\nu(E)$ cumulative number of flares with energy values larger than or equal to $E$ is
\begin{equation}
\nu(E) = c_1 \log E^{-\alpha+1},
\end{equation}
where $c_1$ is a constant number. In logarithmic form it converts to
\begin{equation}
\log \nu(E) = c_2 + \beta \log E,
\end{equation}
i.e., a linear function between $\log \nu$ and $\log E$, where $c_2$ and $\beta=-\alpha+1$ are constant numbers of the linear function, that is the intercept and the slope $1-\alpha$, respectively. 

It has been learned that the low energy turnover of the flare frequency distribution is most probably the result of the detection threshold \citep[cf. e.g.][]{2014ApJ...797..121H}, i.e. the low signal-to-noise ratio of small flares. Therefore we ran flare injection-recovery tests using the code \texttt{allesfitter} \citep{allesfitter-code,allesfitter-paper} to map out the detection bias in the low energy regime. First we set a suitable grid of artificial flares over the FWHMs and amplitudes of the flares, since these two properties feature the relative flare energy. The FWHM-amplitude grid was chosen to cover the relative flare energy range between $t_{\rm eq}$=0.0864-86.4\,sec, convertible to $\log E_f$=33.78-36.78\,[erg] total logarithmic flare energy (see Eq.~\ref{eqen}). To make up the model flare light curves \texttt{allesfitter} adopts the empirical flare template described in \citet{2014ApJ...797..122D}.

\begin{figure}[t!!!!]
   \centering
   \includegraphics[width=\columnwidth]{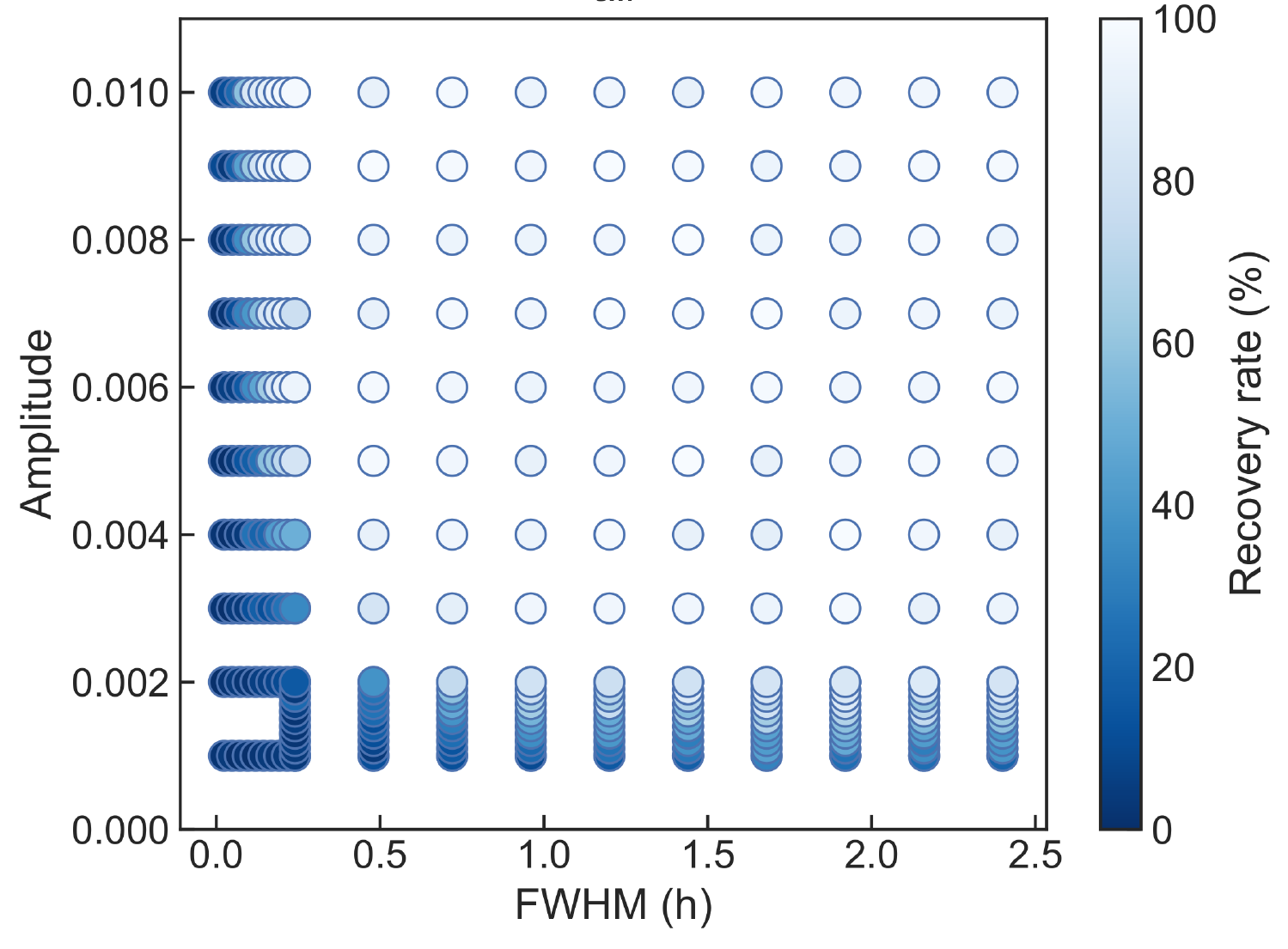}
   \caption{Result of the injection-recovery test. Each circle on the plot represents a given model flare characterized by its FWHM and amplitude. The recovery rate of the model flares is represented by a blue gradient bar where the darker the shade the lower the recovery rate. The energy range in total logarithmic flare energy extends from $\log E$=33.78\,[erg] up to 36.78\,[erg] where the recovery rate virtually reaches 100\%.}
              \label{recovery}
    \end{figure}

The artificial flares were injected into the original \emph{Kepler} light curve and then the flare detection algorithm (described in Sect.~\ref{method}) was applied to recover them, this way characterizing statistically the recovery rate. The resulting injection-recovery plot is seen in Fig.~\ref{recovery} where blue gradient is used to visualize the recovery rate as the function of FWHM and amplitude (the darker the shade the lower the recovery rate). From the test we estimate a detection limit $t_{\rm eq}$ between 5-10\,sec, in agreement with the results in Table~\ref{tab2}. Towards higher FWHM values and amplitudes (i.e., higher energies) the recovery rate increases until the upper right corner of the grid, where the recovery rate virtually reaches $\approx$100\%.
From the recovery rates we derive correction factors (multiplicative inverses) to estimate the real flare numbers at the low energy range. These estimations are used to plot the detection bias-corrected cumulative flare frequency diagram.

   \begin{figure}[thb!!!!]
   \centering
   \includegraphics[width=\columnwidth]{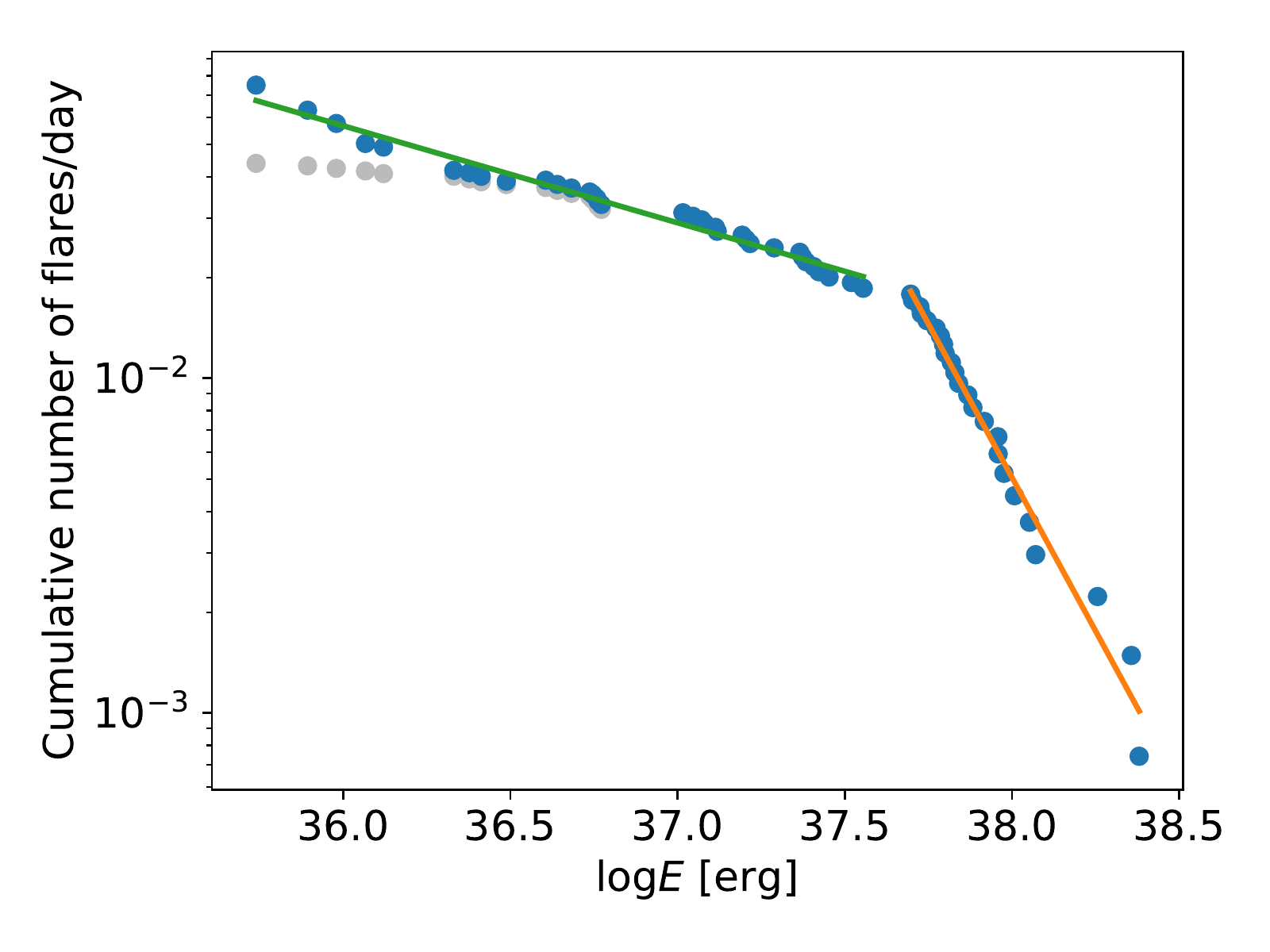}
   \caption{Detection bias-corrected cumulative flare-frequency diagram for KIC\,2852961. The detection-biased original datapoints are plotted in grey colour, above them are the corrected points in blue. The fit to the lower energy range below the breakpoint (green line) yields $\alpha$=1.29$\pm$0.02 parameter, while  the fit for the high energy range above the breakpoint (orange line) gives $\alpha$=2.84$\pm$0.06.}
              \label{corr_cum}
    \end{figure}

The log-log representation of the flare frequency distribution diagram is shown in Fig.~\ref{corr_cum}. At low energy the deviation from linear is nicely reduced compared to the grey dots which indicate the original (uncorrected) frequencies. But more markedly, having a breakpoint at around $E\ge5\times10^{37}$\,erg the distribution deviates from linear exhibiting a slight slope in the low energy region and a much steeper part at the high energy end. Fitting the lower energy range (see the green line in Fig.~\ref{corr_cum}) yields $\alpha=1.29\pm0.02$. (We note that a fit to the uncorrected distribution over the same energy range would yield $\alpha=1.21\pm0.02$.) Applying another fit to the high energy range above the breakpoint at $E\ge5\times10^{37}$\,erg gives $\alpha=2.84\pm0.06$ (see the orange line in Fig.~\ref{corr_cum}). This latter is significantly higher than the fit to the low energy part and higher than that usually derived for flaring dwarf stars \citep[e.g.,][etc.]{Howard_2018,2018ApJ...858...55P,2019A&A...622A.133I,2019ApJS..241...29Y}.
For further discussions on the broken flare energy distribution diagram see Sect.~\ref{disc_cum}.

\section{Correlation between spot modulation and flares}\label{tempdist}

\begin{figure}[t!!!!]
   \centering
   \includegraphics[width=\columnwidth]{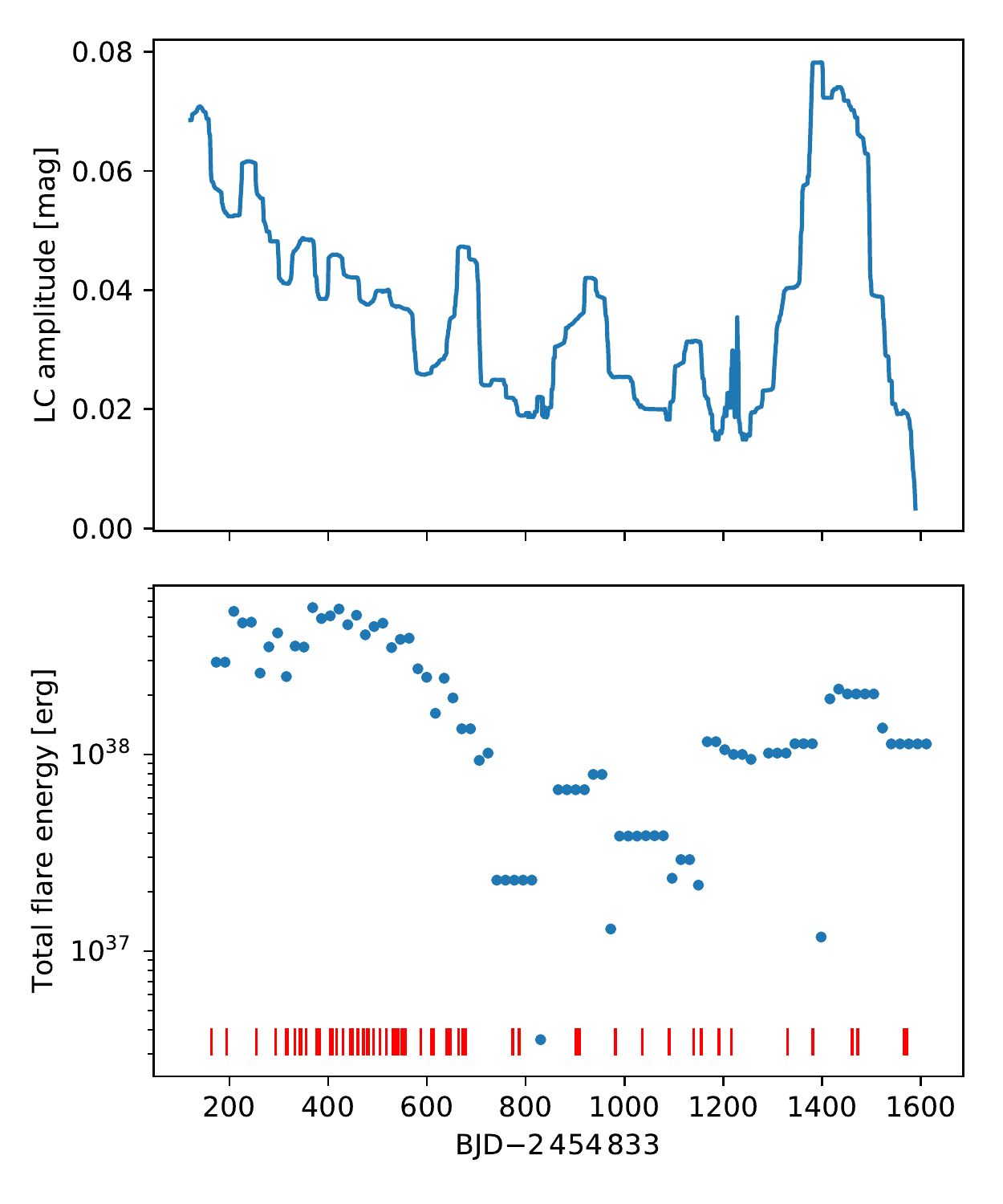}
   \caption{Top: moving average of the amplitude of rotational modulation in the \emph{Kepler} light curve of KIC\,2852961 shown in Fig.~\ref{fig3}. Bottom: the total flare energy within the boxcar of $3P_{\rm rot}$ used for the moving average. The red tick marks below indicate the peak times of the flare events.
   See the text for details.}
              \label{ampenergy}
    \end{figure}

The light curve of KIC\,2852961 in Fig.~\ref{fig3} shows significant change in the amplitude variation and the temporal distribution of flare occurrences during the whole mission. It is clearly seen that in the first third of the mission KIC\,2852961 performed large amplitude rotational modulation with a lot of energetic flares; in the middle of the term the amplitude was getting smaller and the flares were getting less powerful and occurred less frequently; at the end of the term the amplitude increased again together with the average flare energies, although the flare frequency did not change much during the second half of the observing term. To quantify this observation, in Fig.~\ref{ampenergy} we plot the simple moving average of the overall amplitude change of the light curve cleaned from flares (top panel) together with the total flare energy within the same boxcar, which was set to be $3P_{\rm rot}$ (bottom panel). In the lower edge of the bottom panel we mark the individual flare events by red ticks. The plot indicates that there is indeed a connection between the rotation amplitude (as an indicator of magnetic activity) and the overall magnetic energy released by flares. Especially interesting is the second half of the observing term starting with small amplitudes at BJD$\approx$800 (+2454833) with a few smaller flares. After a few rotation periods, at around BJD$\approx$900 the amplitude started to increase significantly, just like the flare energies. This pattern is even more apparent from BJD$\approx$1300, where the amplitude of the rotational modulation increases more rapidly which coincides with the overall flare energy increase (without the increase of the average flare count). This result suggests a general scaling effect behind the production flares in the sense that there are more and/or more energetic flares when having more/larger active regions on the stellar surface and the flare activity is lower when there are less/smaller active regions. In Sect.~\ref{disc3} we give a few more words on this topic, while in Appendix~\ref{A1} we demonstrate the reliability of the observed PDCSAP amplitudes shown in Fig.~\ref{fig3}.

\section{Discussions}\label{disc}

\subsection{On the stellar evolution, rotation and differential rotation}\label{disc1}

The revised stellar parameters of KIC\,2852961 listed in Table~\ref{tab1} are in agreement with the new \emph{Gaia} DR-2 observations and are more consistent with each other. The mass of about 1.7\,$M_{\odot}$ is considerably higher than the formerly suggested mass of $\approx$0.9\,$M_{\odot}$ (NASA Exoplanet Archive). The higher mass we find is suggestive of a faster evolution or younger age of $\approx$1.7Gyr as well (which is tightly interrelated with mass). Taking the relation $t_{\rm MS}\propto M^{-2.5}$ between the main sequence period and mass with 0.9\,$M_{\odot}$ instead of 1.7\,$M_{\odot}$ and supposing no significant mass loss at the red giant branch \citep[cf. e.g.,][]{2015MNRAS.448..502M} would mean a much slower evolution on the main sequence of $t_{\rm MS}\approx$13\,Gyr, which would be unreasonably long.
Using the mass-radius relation \citep[see e.g.,][]{1981gask.book.....M,1991Ap&SS.181..313D} one can estimate $R_{\rm MS}\approx2.0\,R_{\odot}$ for the main sequence, i.e., a typical A5-F0 type progenitor \citep[cf.][]{2012ApJ...746..101B,2013ApJ...771...40B}.
Assuming that our SB1 target \citep[cf.][]{2020arXiv200413792G} has a distant and/or too low mass companion with having insignificant influence and the angular momentum is conserved over the red giant branch, applying the relation $P_{\rm rot}\propto R^2$ will give $P_{\rm rot,MS}$ of $\approx$1\,d period at the end of the main sequence. This rotation rate is high, but not unusual for an effectively single A5-F0 star just leaving the main sequence, since in stars with >1.3\,$M_{\odot}$ mass the lack of deep convective envelope does not enable to generate enough strong magnetic fields to maintain effective magnetic braking over the main-sequence \citep[][]{2013ApJ...776...67V}. Still, other mechanisms could also be considered to spin up the surface of an evolved star on the red giant branch. A certain fraction of red giants have undergone such spin up phases \citep[e.g.,][]{2011ApJ...732...39C,2017A&A...605A.111C}, which may involve mixing processes (\citealt[e.g.,][]{1989ApJ...346..303S}, but see also \citealt{2014A&A...571A..74K,2017A&A...606A..42K}), planet engulfment \citep[][]{1999MNRAS.308.1133S,2016A&A...596A..53K}, binary mergers  \citep[][]{1976ApJ...209..829W,1998A&A...336..587S} or other, less known mechanisms.
However, the lack of systematic radial velocity measurements does not enable to know whether KIC\,2852961 is a member of a wide binary system or rather a close binary. In the latter case the stellar rotation is probably synchronized to the orbital motion which would evidently account for the 35.5\,day period.

Interpreting the multiple peaked power spectrum in Fig.~\ref{fig4} together with the different photometric periods obtained for different datasets (see Sect.~\ref{intro}) as the signs of surface differential rotation, we give a low end estimate for the surface shear parameter $\Delta P/P$ to be $\approx$0.1. This value agrees with the result in \citet[][see their Fig.~1]{2017AN....338..903K}, where authors predict $\approx$0.17 for a single giant rotating at a similar rate, based on an empirical relationship between rotation and differential rotation. From photometric observations only, however, usually it is not evident to determine, whether the differential rotation is solar-type (i.e., when the angular velocity has its maximum at the equator and decreases with latitude) or oppositely, antisolar \citep[but see][]{2015A&A...576A..15R}.

\subsection{On the cumulative flare frequency diagram}\label{disc_cum}

With the detection bias-correction even the broken flare energy distribution is more apparent.
Such a broken distribution has already been observed in a few dwarf stars 
\citep{1989SoPh..121..375S,2018ApJ...858...55P}
including the Sun \citep{2003A&AT...22..325K}. \citet{2018ApJ...854...14M} interpreted this feature as energy release from twisted magnetic loops at different energies: below and above a critical energy when the loop size becomes higher than the local scale height depending on the local field strength and density. Below and above the critical energy the power law slopes are different and the breakpoint is at the critical energy. This critical energy is different from star to star, and the flare energy distribution does not necessarily contain it, therefore - apart from the natural undersampling at low energies - a single power law can also describe flare energy distributions of many stars. This scenario of  \citet{2018ApJ...854...14M} was developed for dwarf stars taking into account only simple flares. Specifically, \citet{2013ApJS..209....5S} derived $\alpha = 2.0-2.2$ power-law indices for {\it Kepler} G-dwarf superflare stars while $\alpha$=1.53 for the \say{normal} solar flares. From \emph{Kepler} short-cadence data \citet{2015EP&S...67...59M} detected 187 flares on 23 solar-like stars and derived $\alpha$=1.5 between the $10^{33}$$-$$10^{36}$\,erg flare energy range. \citet{2018ApJ...854...14M} suggested a critical energy around $10^{32}$$-$$10^{33}$\,erg for solar-like stars.
Our result of KIC2852961 is $\alpha = 2.84$ and 1.29 for the superflare and normal flare part of the frequency distribution, respectively, with the critical energy being about $10^{37.6}$\,erg. The big difference between the critical energies and the slopes of the distributions are very probably due to the differences between the atmospheric parameters (e.g., $\log g$, density) of dwarf and giant stars and the characteristic magnetic field strengths. Anyhow, the aforesaid interpretation of the broken distribution supports the idea that flares/superflares in solar-type dwarf stars and in flaring giants have common origin but reveal themselves on different energy scales.

But further explanations may also arise. \citet{2010ApJ...710.1324W} studied a sample of small X-ray flares observed by \emph{GOES} satellite, all erupted form one active region on the Sun. In the flare frequency distribution they found a departure from the standard power-law ($1.88\pm0.12$) which was interpreted as a possible result of finite magnetic free energy for flaring.
\citet{2019A&A...622A.133I} discussed three possible reasons behind the broken power-laws: {\it i)} undetected multiplicity of flaring stars with different flare frequencies superimposed; {\it ii)} flares can be produced by different active regions on the same star having different flare statistics; {\it iii)} a close-in planetary companion could trigger flares with a different mechanism adding events to the intrinsic flare distribution.
Finally, the work by \citet{2006ApJ...650L.143Y} could bring an additional perspective regarding the different power-law indices: authors found that the power-law indices for solar flares without coronal mass ejections (CMEs) are steeper than those for flares with CMEs. This interpretation might also work for stellar flares, however, so far only a handful of stellar CMEs has been detected, all of them by spectroscopy \citep[see the recent statistical study by][and their references]{2020MNRAS.493.4570L}. Accordingly, without spectroscopic observations it is not possible to draw such a distinction in our flare sample.

\begin{figure}[thb]
   \centering
   \includegraphics[width=1.0\columnwidth]{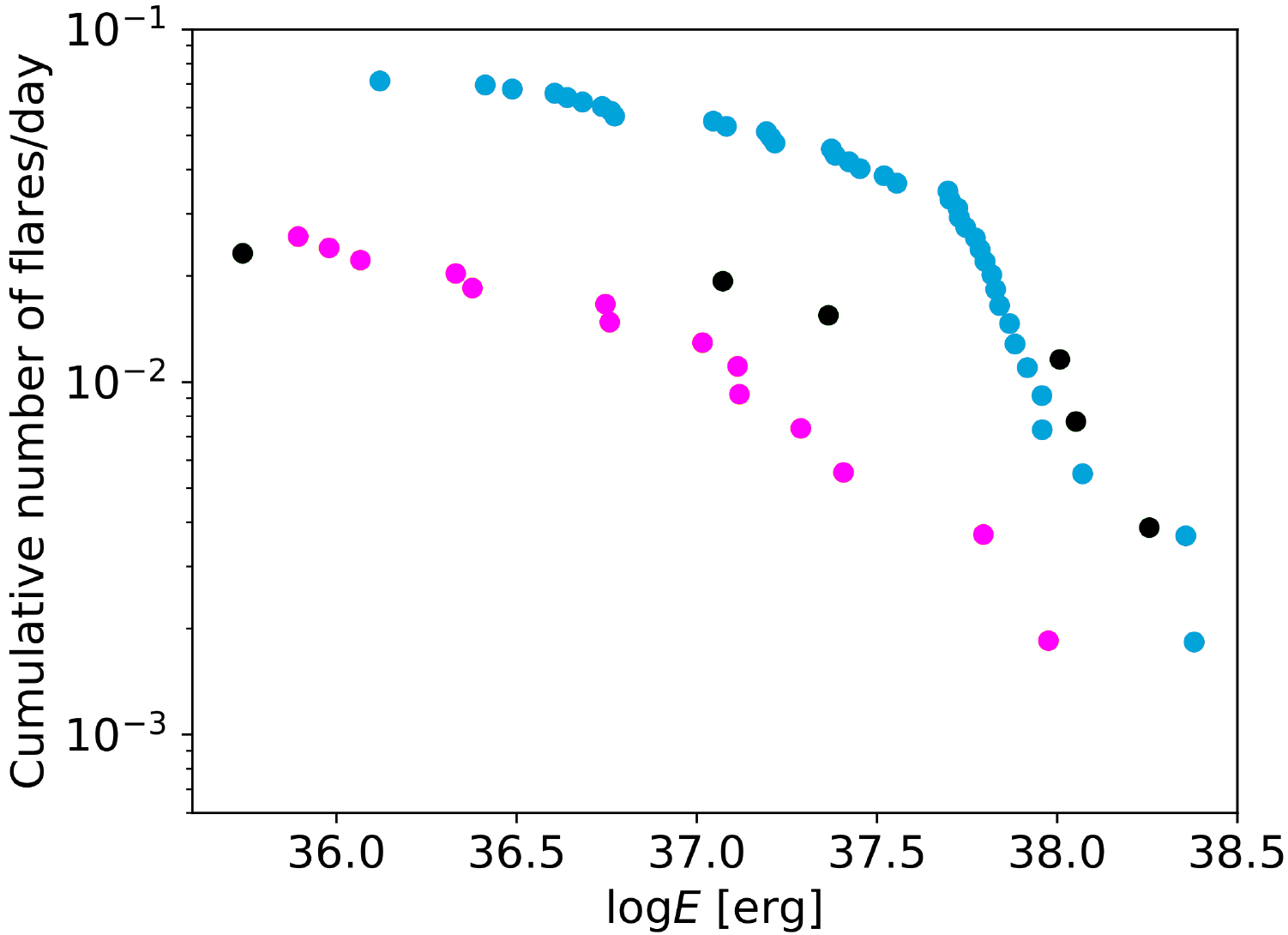}
   \caption{Cumulative  flare-frequency diagram for three different phases of the \emph{Kepler} light curve of KIC\,2852961. Blue dots correspond to the FFD for the first period of the \emph{Kepler} data until BJD=700 (+2454833), when the rotational modulation showed large amplitudes. The FFD plotted in magenta represents the small amplitude period between BJD=700-1300, while the third FFD in black corresponds to the third part of the light curve after BJD=1300, when the amplitude increased again. See the text for explanation.}
              \label{fig_cum3}
    \end{figure}

\citet[][see their Figs.~6-8]{2017PASJ...69...41M} demonstrated how the flare-frequency distribution changes with spot sizes on the Sun and stars: with increasing spot area increasing flare-frequency was found at a certain energy level. In our case the starspot area is supposed to change with the light curve amplitude (cf. Sect.~\ref{tempdist}), therefore similar temporal change in the flare frequency distribution is expected when comparing different parts of the \emph{Kepler} light curve. We cut the whole light curve into three parts: the first part, referred as the \say{first maximum phase} is until BJD=700 (+2454833), the second \say{minimum phase} is limited from BJD=700 to 1300, which is followed by the \say{second maximum phase} from BJD=1300 until the end of the dataset (cf. Fig.~\ref{fig3} and Fig.~\ref{ampenergy}). After this, flare-frequency diagram was derived for each phase separately and the diagrams are plotted together in Fig.~\ref{fig_cum3}. Clearly, compared to the first maximum phase (blue dots in the figure), in the minimum phase (plotted in magenta), when the starpot area decreased, the flare frequency at a given energy level is decreased as well. And despite the small flare count during the third phase after BJD=1300 (second maximum phase), with increasing amplitude the corresponding FFD  (black dots in Fig.~\ref{fig_cum3}) shows increasing flare frequency again. Our result definitely supports the finding in
\citet[][]{2017PASJ...69...41M}.

\subsection{On the flare occurrence along the rotational phase}\label{flareoccur}

\begin{figure}[tb]
   \centering
   \includegraphics[width=\columnwidth]{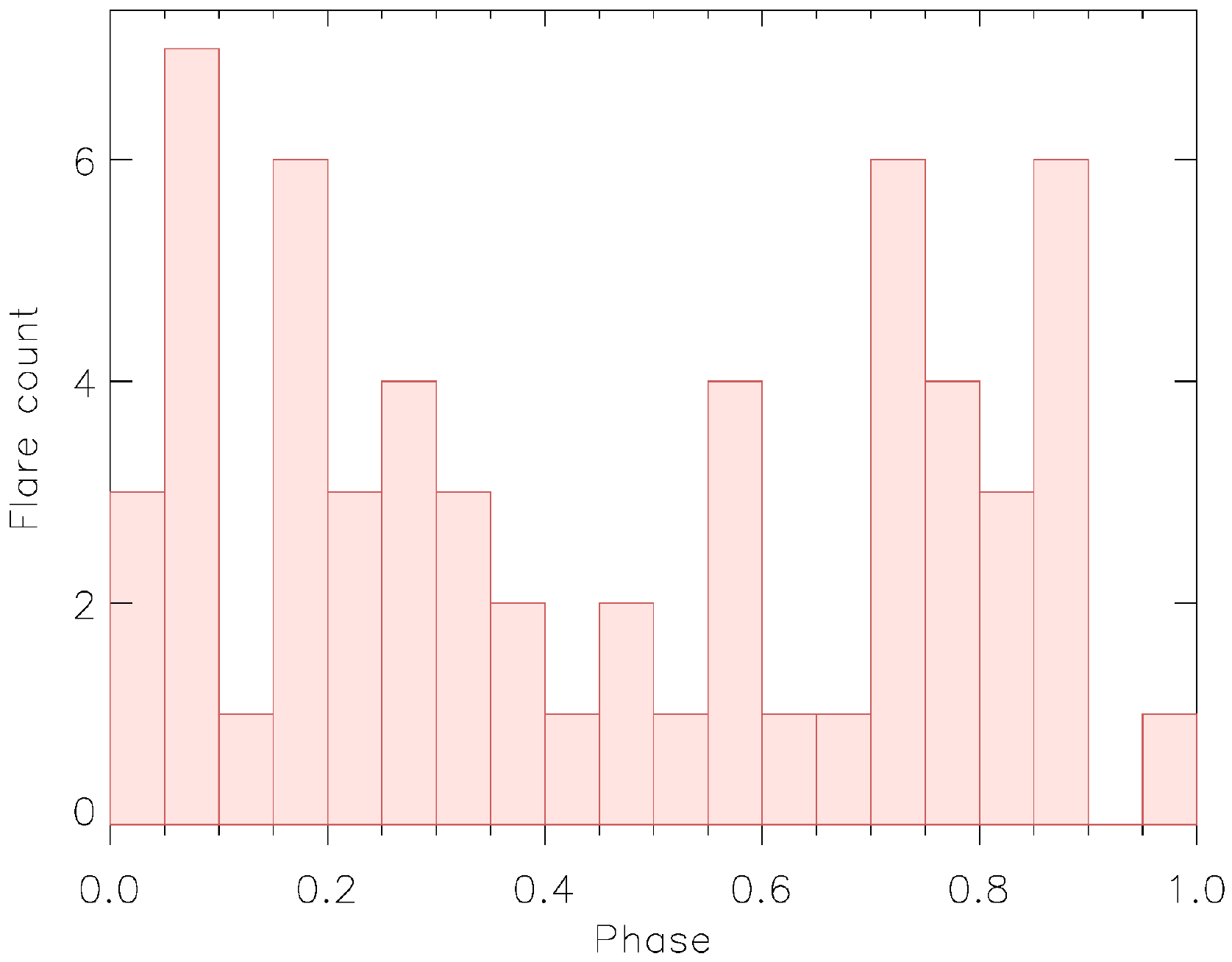}
   \caption{Flare occurence over the rotational phase. The plot indicates no preferred phase for flares.}
              \label{fig_flarehisto_phase}
    \end{figure}

Since spots and flares are related phenomena, therefore, flares expectedly occur when the spotted hemisphere of the star turns in view. In other words, over the rotational phase more flares occur when more/larger spots are in the apparent stellar hemisphere, i.e. at around brightness minima \citep[e.g.][]{1994A&A...287..575C,2000A&A...357..608S}. However, in the bottom panel of Fig.~\ref{fig4} flares seem to occur randomly all over the rotational phase. In Fig.~\ref{fig_flarehisto_phase} we plotted the rotational phase distribution of the 59 flares. The plot indicates that there is no preferred phases for flare occurrence. According to our K$-$S test the phase distribution of flares is random at 97\% significance level. Such a result may be surprising in comparison with the solar case \citep[cf.][]{2014MNRAS.441.2208G}, but it is not unusual among flaring stars: recent statistical analyses for M dwarf stars \citep{2018MNRAS.480.2153D,2019MNRAS.489..437D} have revealed no indication that flares would occur more frequently at rotational phases around brightness minima. This can partly be understood when considering large starspots (active regions) which cover much bigger fraction of the visible stellar disk compared to sunspots. On stars, flare loops interconnecting distant magnetic regions can be more extended, that is, more visible as well, compared to solar flare loops over bipolar regions. Also, when the inclination angle is not very high, flares could come from spots near the visible pole over the entire rotational phase. For other alternatives see the discussion in \citet{2018MNRAS.480.2153D}.

\subsection{On the flare energy range, loop size and statistics}\label{disc2}

\citet{2019ApJ...876...58N} discussed on the maximum magnetic energy $E_{\rm mag}$ available in a spotted star, which eventually may be converted to flare energy to a certain extent. Here we follow their method to estimate  $E_{\rm mag}$ on KIC\,2852961. When assuming an appropriate spot temperature, the overall spot coverage of the star (leastwise the spot coverage difference between the maximum and minimum rotational phases) can be estimated from the relative amplitude of stellar brightness variation. Taking $T_{\rm spot}=3500$\,K \citep[cf.][]{2005LRSP....2....8B} and $\Delta F/F\approx0.04$ maximum relative amplitude from the \emph{Kepler} light curve plotted in Fig.~\ref{fig3} yields a maximum spot coverage (\say{spot size}) of $L\approx0.24 R_{\star}$ \citep[see Eq.~3 in][]{2019ApJ...876...58N}. Taking $L$ and assuming $B=3.0$\,kG as a reasonable value of the average magnetic flux density in the spot \citep[cf.][]{2019ApJ...876...58N}, according to the equation in \citet[][see their Eq.~1]{2011LRSP....8....6S} the maximum available magnetic energy is estimated to be at least $E_{\rm mag}\approx3.5\times10^{39}$\,erg. However, only a small part of this energy is available to feed a flare because it is distributed as potential field energy. Nevertheless, being $\approx$15 times more than the highest flare energies in Table~\ref{tab2}, this rough estimation is in fair agreement with the observed flare energies of KIC\,2852961.

As has been mentioned in Sect.~\ref{flareoccur}, flare loops interconnecting large and distant spotted areas on magnetically active giant stars can be quite extended. A good example is the active giant $\sigma$\,Gem which is very similar to KIC\,2852961, with a rotational period of 19.6 days, $R_\star$=12.3\,$R_{\odot}$ and  $T_{\rm eff}$=4630\,K \citep{2001A&A...373..199K}. 
Using \emph{Extreme Ultraviolet Explorer} observations \citet{2006ApJS..164..173M} studied the lengths of flaring loops in different types of active stars.
Analysing flares on $\sigma$\,Gem they found a loop length of 0.85\,$R_*$ and 80\,G as minimum coronal magnetic field strength for a flare lasting 22 hours. Starting from this, we may adopt $\approx$0.5$R_\star$ as a characteristic flare loop size and assume 100\,G as minimum coronal magnetic field strength \citep[cf.][]{2006ApJS..164..173M}. According to \citet[][see their Eq.~1]{2017ApJ...851...91N}, this would yield at least $E_{\rm mag}\approx3\times10^{39}$\,erg potential energy, i.e., very similar to the value derived from spot coverage and light curve amplitude in the previous paragraph. This finding suggests that flare loop sizes for the largest flares of KIC\,2852961 are indeed in the order of $R_\star$ (but of course, could be quite different from flare to flare).

   \begin{figure}[ttt!!!!]
   \centering
   \includegraphics[width=0.94\columnwidth]{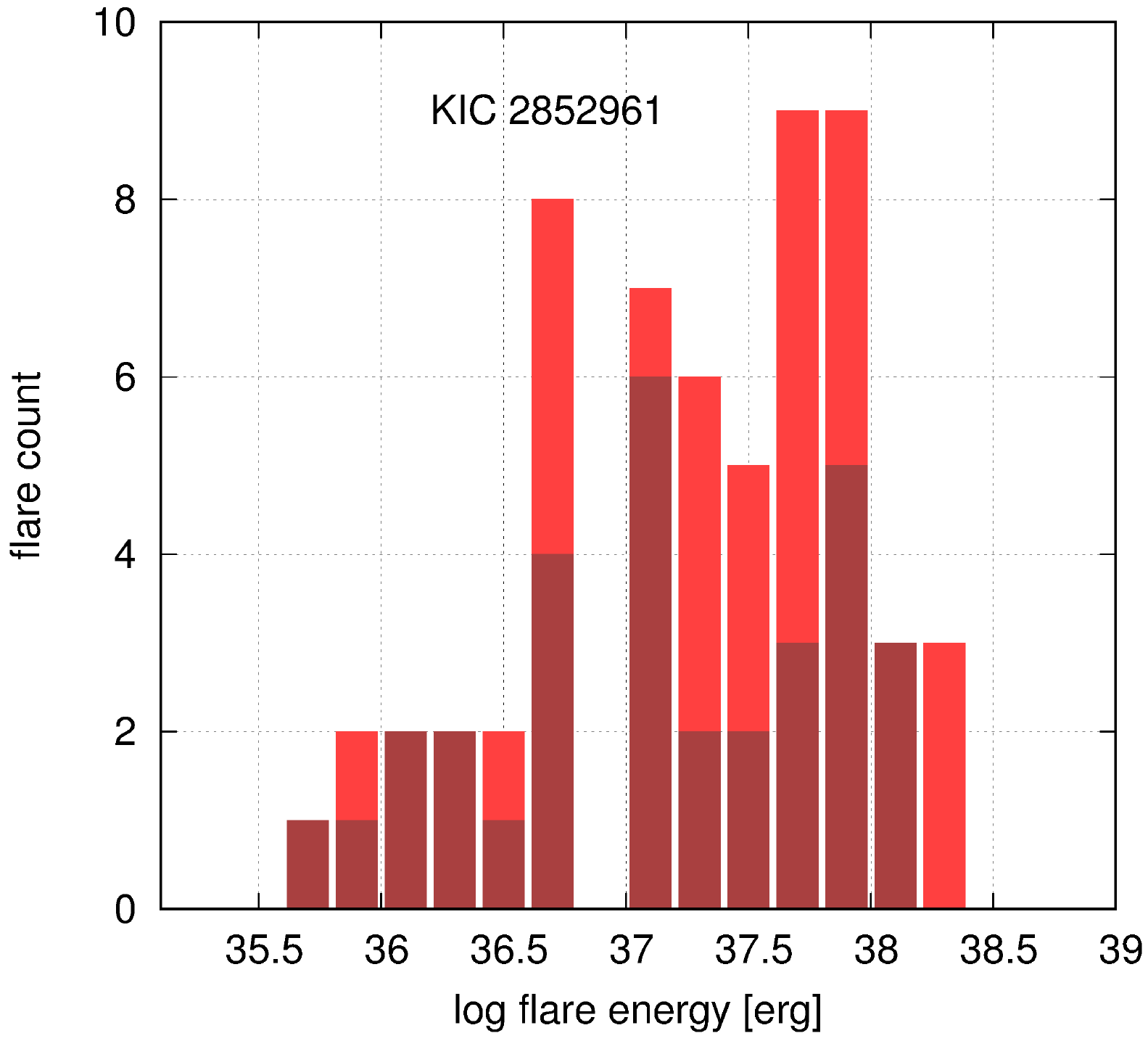}
   \caption{Flare statistics of KIC\,2852961. The fraction of the 32 regular flares in the full sample of 59 flares is plotted in dark red. See the text for details.}
              \label{histograms}
    \end{figure}

Comparing the flare energies of a mixed sample of flares observed on giant stars by \emph{Kepler} \citep{2019ApJS..241...29Y} against the flare energies of KIC\,2852961 we find that KIC\,2852961 flares are more powerful. Both the 35.503 mean and the 35.417 median of the $\log E_f$ values in the cited sample of 6842 flares from giant stars are well below our values of 37.286 and 37.384, respectively. Nevertheless, our flare sample is statistically much smaller.

The energy distribution function of KIC\,2852961 is plotted in Fig.~\ref{histograms}. The histogram exhibits relatively more flares at high energies.
Such an excess, however, could be a bias from accounting the complex (and so possibly simultaneous) flares as single ones. Therefore, to see the energy distribution of the regular flares separately, we clean the flare sample by excluding the complex events (27 of 59 in Table~\ref{tab2}). The histogram of the fraction of regular flares is overplotted in Fig.~\ref{histograms} (see in dark red) showing similar increase at high energies. This supports that the high energy excess in the histogram is not simply an artifact; it seems as if KIC\,2852961 may favor the production of superflares. But again, our sample of altogether 59 flares (with 32 regular shaped among them) is statistically scant, which should also be borne in mind.

\subsection{On the scaling effect behind different flare statistics}\label{disc3}

\citet{2019A&A...622A.133I} published average slopes of frequency distribution of flare stars in three open clusters (Pleiades, Praesepe and M67) of different ages (0.125, 0.63 and 4.3\,Gyr, respectively) observed by \emph{Kepler}. The power-law exponents were found not to change with age, and the same was found by \citet{2019ApJ...871..241D}, reflecting a probable universal flare producing mechanism, regardless of age (+metallicity, rotational evolution, etc.).

\begin{figure}[tt!!!!]
   \centering
   \includegraphics[width=\columnwidth]{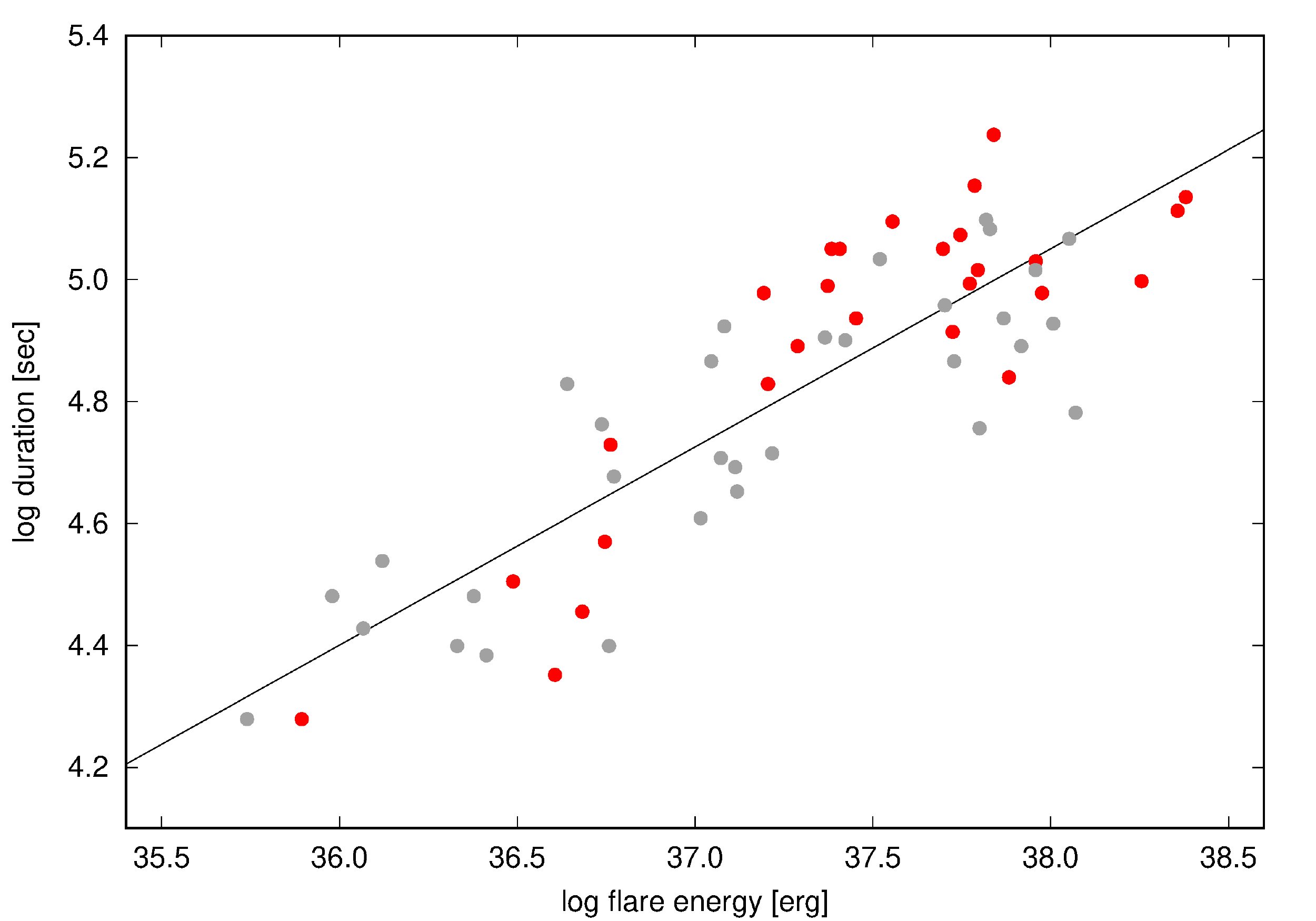}
   \caption{Flare duration vs. flare energy for KIC\,2852961 in log-log interpretation. Grey dots are regular flares while red dots indicate the complex events.
   The fit supports a power-law relationship with $0.325(\pm0.026)$ exponent.}
              \label{durations}
    \end{figure}

\citet{2015MNRAS.447.2714B} studied \emph{Kepler} flare stars of different luminosities and found that higher luminosity class stars, including giants, have generally higher energy flares. This finding \citep[][see Fig.~10 in that paper]{2015MNRAS.447.2714B} is attributed by the author to a scaling effect, i.e., in the larger active region of a larger star more energy can be stored from the same magnetic field strength. Another important finding of \citet{2015MNRAS.447.2714B} is that stars with lower surface gravities have longer duration flares. \citet{2015EP&S...67...59M} found that flare duration increases with flare energy, which can be explained by assuming that the time scale of flares emerging from the vicinity of starspots is determined by the characteristic reconnection time \citep[][]{2011LRSP....8....6S}. Accordingly, the relationship between $\Delta t$ and $E_f$ for solar-type main sequence stars is expected to be $\Delta t\propto E_f^{1/3}$. For comparison, in Fig.~\ref{durations} we plot $\log\Delta t$ vs. $\log E_f$ values for KIC\,2852961. The dots are fitted by a power-law in the form of
\begin{equation}
\log \Delta t {\rm [sec]}= -7.30(\pm0.96) + 0.325(\pm0.026)\log E_f {\rm [erg]}.
\end{equation}
We note that fitting either the regular flares (grey dots) only or the complex events (red dots) would yield 0.287$\pm$0.032 and 0.361$\pm$0.042 slopes, respectively. However, due to small sample sizes, the difference is statistically not significant.
The slope of 0.325$\pm$0.026 in Fig.~\ref{durations} is similar to the value of 0.39$\pm$0.026 in \citet{2015EP&S...67...59M} obtained for G-type main sequence stars, supporting the idea that the differences between flare energies are due to size effect \citep[cf.][]{2015MNRAS.447.2714B}. This scaling idea is echoed by the avalanche models for solar flares, which regard flares as avalanches of many small reconnection events \citep[for an overview see][]{2001SoPh..203..321C}. This statistical approach could consistently be extended to provide a common framework for solar flares and stellar superflares over many orders of magnitude in energy.

The correlation between spot modulation and flare occurrence presented in Sect.~\ref{tempdist} suggests a clear connection between the level of spot activity and the overall flare energy. \citet{2018ApJS..236....7H} investigated the relationship between the magnetic feature activity and flare activity of three solar-type stars observed by \emph{Kepler} and concluded that both magnetic feature activity and flare activity are influenced by the same source of magnetic energy, i.e. the magnetic dynamo, similarly to the solar case \citep[][]{2015LRSP...12....4H}. These all are concordant with the scaling idea by \citet{2015MNRAS.447.2714B}, i.e. when having larger active regions more magnetic energy can be stored and thus released by flares \citep[but see also][for a common dynamo scaling in all late-type stars]{2020NatAs.tmp...46L}.

\section{Summary and conclusion}\label{summary}

In this paper the flare activity of the late G-early K giant KIC\,2852961 
was analyzed using the full \emph{Kepler} data (Q0-Q17) and one \emph{TESS} light curve (Sector 14 from July-August 2019) in order to study the flare occurrence and other signs of magnetic activity. Foremost, adopting the \emph{Gaia} DR-2 parallax we revised the astrophysical data of the star and more reliable parameters were derived, in agreement with the position in the H-R diagram. We found altogether 59 flare events in the \emph{Kepler} time series and another one in the much shorter \emph{TESS} light curve. Logarithmic flare energies range between 35.74--38.38\,[erg], i.e., almost three orders of magnitude, however, the detection cutoff at low energy end is very likely due to the noise limit. We derived a cumulative flare frequency distribution diagram which deviated from a simple power-law in the sense that different exponents were obtained for the lower and the higher energy ranges, having a breakpoint between them. We reviewed a couple of possible explanations to understand the broken power-law. Flare counts and total flare energies per unit time show temporal variations, which are related to the average light curve amplitude. A straightforward interpretation is that the higher the level of spot activity the more the overall magnetic energy released by flares and/or superflares. This exciting result supports the assumption that differences in flare (superflare) energies of different luminosity class targets are very likely due to size effect.

\begin{acknowledgements}
Authors are grateful to the anonymous referee for his/her valuable comments which helped to improve the manuscript. 
      Authors thank Andrew Vanderburg at Harvard Smithsonian Center for Astrophysics for his help in understanding the action of Presearch Data Conditioning (PDC) module in \emph{Kepler} data processing. This work was supported by the Hungarian National Research, Development and Innovation Office grant OTKA K131508, KH-130526
      and by the Lend\"ulet Program  of the Hungarian Academy of Sciences, project No. LP2018-7/2019. Authors from Konkoly Observatory acknowledge the financial support  of the Austrian-Hungarian  Action  Foundation (95 \"ou3, 98\"ou5, 101\"ou13). M.N.G. acknowledges support from MIT’s Kavli Institute as a Torres postdoctoral fellow.
      Data presented in this paper are based on observations obtained with the Hungarian-made Automated Telescope Network, with stations at the Submillimeter Array of the Smithsonian Astrophysical Observatory (SAO), and at the Fred Lawrence Whipple Observatory of SAO. IRAF used in this work was distributed by the National Optical Astronomy Observatory, which was managed by the Association of Universities for Research in Astronomy (AURA) under a cooperative agreement with the National Science Foundation.
      This research has made use of the NASA Exoplanet Archive, which is operated by the California Institute of Technology, under contract with the National Aeronautics and Space Administration under the Exoplanet Exploration Program. This work presents results from the European Space Agency (ESA) space mission Gaia. Gaia data are being processed by the Gaia Data Processing and Analysis Consortium (DPAC). Funding for the DPAC is provided by national institutions, in particular the institutions participating in the Gaia MultiLateral Agreement (MLA). The Gaia mission website is \texttt{https://www.cosmos.esa.int/gaia}. The Gaia archive website is \texttt{https://archives.esac.esa.int/gaia}. 
\end{acknowledgements}

\bibliographystyle{aa}
\bibliography{kic2852961}

%

\begin{appendix}
\section{Reliability of the rotational amplitudes}\label{A1}
\begin{figure}[thb]
\includegraphics[width=\columnwidth]{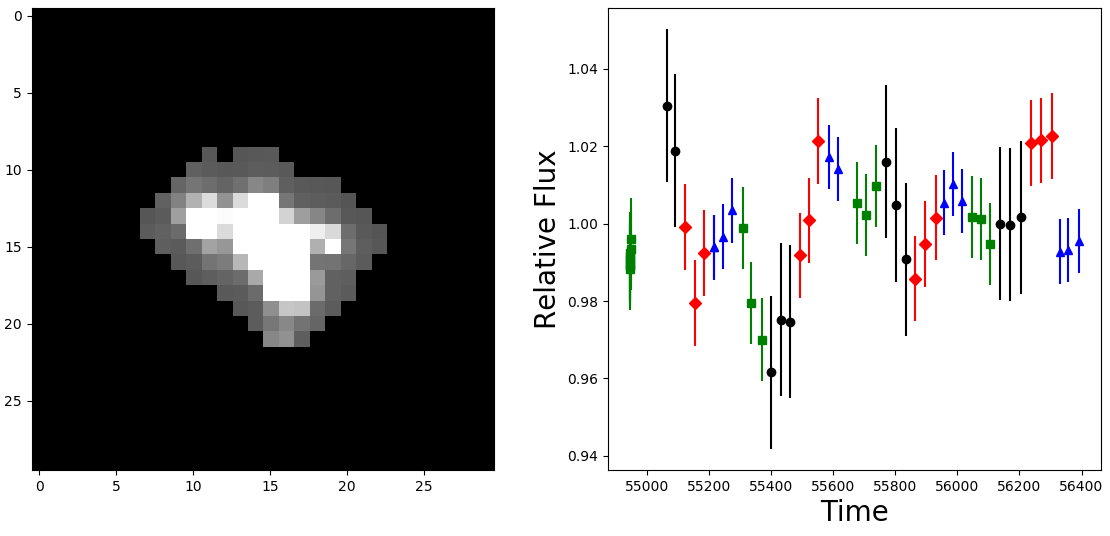}
\caption{\emph{Kepler} FFI photometry of KIC\,2852961 obtained from using {\it f3} code by \citet{2017ApJ...851..116M}. In the left panel the applied original aperture is seen, while in the right panel the FFI light curve is plotted. Quarterly changing colours indicate four different positions of the rotated spacecraft (i.e. the CCD). Plotted errors are supposedly instrumental.}
\label{f3a}
\end{figure}

The 30-minute (long cadence) \emph{Kepler} photometry is ideal for characterizing short-term signals such like planetary transits. However, on longer terms of $\approx$50 days or more, such brightness variations like long-term magnetic activity changes were removed by the processing pipeline. Moreover, \emph{Kepler} data can also be affected by quarterly systematics as the instrument was rotated by 90 degrees in every three months. Therefore long-term \emph{Kepler} data may either obscure true activity related variability or, on the contrary, mimic false astrophysical signals \citep[see e.g.,][]{2012PASP..124..963K}. But long-term variability can be recovered from the full frame images (FFIs), which were recorded monthly during the primary mission. To recover the long-term brightness variability of KIC\,2852961 and to rule out possible unreliable changes we processed the available FFIs by following the method described in \citet{2017ApJ...851..116M}. We use the open source code {\it f3} provided in the paper, which carries out photometry from the FFIs by centering an appropriate aperture on the given target on every full frame image over the primary \emph{Kepler} mission. The resulting long-term FFI photometry of KIC\,2852961 is plotted in Fig.~\ref{f3a}. The plot reflects brightness changes mostly related to the amplitude changes already seen in Fig.~\ref{fig3}. However, the monthly sampling of the FFIs is quite close to the 35.5\,day rotation period of the star, which is also markable in the track of the light curve: the frequency of $\approx$1/200\,d$^{-1}$ seen in the FFI light curve in Fig.~\ref{f3a} is just an $f_a$ alias frequency obtained as the difference between the sampling and data frequencies, i.e. $f_a=|f_d-f_s|$. Still, the FFI light curve amplitudes are in the order of the light curve amplitudes in Fig.~\ref{fig3} obtained by Presearch Data Conditioning (PDC) with Simple Aperture Photometry (SAP), which confirms their reliability.

On the other hand, beside the rotation amplitude related variability, no other significant change can be seen over the observing run, in other words, during the four years of the primary \emph{Kepler} mission no long-term variability in the overall (average) brightness level of KIC\,2852961 can be verified.

\end{appendix}
\end{document}